\documentclass[letter,prd,aps,twocolumn,floatfix,superscriptaddress]{revtex4}

\usepackage{amssymb,amsmath,graphicx}
\usepackage{hyperref}
\usepackage[usenames,dvipsnames]{color}
\usepackage{array}
\makeatletter
\renewcommand*{\p@section}{\S\,}
\renewcommand*{\p@subsection}{\S\,}

\makeatother

\def\jnl@style{\it}
\def\aaref@jnl#1{{\jnl@style#1}}

\def\aaref@jnl#1{{\jnl@style#1}}

\def\aj{\aaref@jnl{AJ}}                   
\def\araa{\aaref@jnl{ARA\&A}}             
\def\apj{\aaref@jnl{ApJ}}                 
\def\apjl{\aaref@jnl{ApJ}}                
\def\apjs{\aaref@jnl{ApJS}}               
\def\ao{\aaref@jnl{Appl.~Opt.}}           
\def\apss{\aaref@jnl{Ap\&SS}}             
\def\aap{\aaref@jnl{A\&A}}                
\def\aapr{\aaref@jnl{A\&A~Rev.}}          
\def\aaps{\aaref@jnl{A\&AS}}              
\def\azh{\aaref@jnl{AZh}}                 
\def\baas{\aaref@jnl{BAAS}}               
\def\jrasc{\aaref@jnl{JRASC}}             
\def\memras{\aaref@jnl{MmRAS}}            
\def\mnras{\aaref@jnl{MNRAS}}             
\def\pra{\aaref@jnl{Phys.~Rev.~A}}        
\def\prb{\aaref@jnl{Phys.~Rev.~B}}        
\def\prc{\aaref@jnl{Phys.~Rev.~C}}        
\def\prd{\aaref@jnl{Phys.~Rev.~D}}        
\def\pre{\aaref@jnl{Phys.~Rev.~E}}        
\def\prl{\aaref@jnl{Phys.~Rev.~Lett.}}    
\def\pasp{\aaref@jnl{PASP}}               
\def\pasj{\aaref@jnl{PASJ}}               
\def\qjras{\aaref@jnl{QJRAS}}             
\def\skytel{\aaref@jnl{S\&T}}             
\def\solphys{\aaref@jnl{Sol.~Phys.}}      
\def\sovast{\aaref@jnl{Soviet~Ast.}}      
\def\ssr{\aaref@jnl{Space~Sci.~Rev.}}     
\def\zap{\aaref@jnl{ZAp}}                 
\def\nat{\aaref@jnl{Nature}}              
\def\iaucirc{\aaref@jnl{IAU~Circ.}}       
\def\aplett{\aaref@jnl{Astrophys.~Lett.}} 
\def\apspr{\aaref@jnl{Astrophys.~Space~Phys.~Res.}}
\def\bain{\aaref@jnl{Bull.~Astron.~Inst.~Netherlands}} 
\def\fcp{\aaref@jnl{Fund.~Cosmic~Phys.}}  
\def\gca{\aaref@jnl{Geochim.~Cosmochim.~Acta}}   
\def\grl{\aaref@jnl{Geophys.~Res.~Lett.}} 
\def\jcp{\aaref@jnl{J.~Chem.~Phys.}}      
\def\jgr{\aaref@jnl{J.~Geophys.~Res.}}    
\def\jqsrt{\aaref@jnl{J.~Quant.~Spec.~Radiat.~Transf.}}
\def\memsai{\aaref@jnl{Mem.~Soc.~Astron.~Italiana}}
\def\nphysa{\aaref@jnl{Nucl.~Phys.~A}}   
\def\physrep{\aaref@jnl{Phys.~Rep.}}   
\def\physscr{\aaref@jnl{Phys.~Scr}}   
\def\planss{\aaref@jnl{Planet.~Space~Sci.}}   
\def\procspie{\aaref@jnl{Proc.~SPIE}}   

\begin{document}

\date{\today}
\title{Unequal mass binary neutron star mergers and multimessenger signals}

\author{Luis Lehner}
\affiliation{Perimeter Institute for Theoretical Physics,Waterloo, Ontario N2L 2Y5, Canada}
\author{Steven L. Liebling}
\affiliation{Department of Physics, Long Island University, Brookville, New York 11548, USA}
\author{Carlos Palenzuela}
\affiliation{Departament de F\'isica, Universitat de les Illes Balears and Institut d'Estudis Espacials e Catalunya, Palma de Mallorca, Baleares E-07122, Spain}
\author{O. L. Caballero}
\affiliation{Department of Physics, University of Guelph, Guelph, Ontario N1G 2W1, Canada}
\author{Evan O'Connor}
\affiliation{Department of Physics, North Carolina State University, Raleigh, North Carolina 27695, USA}
\author{Matthew Anderson}
\affiliation{Pervasive Technology Institute, Indiana University, Bloomington, IN 47405, USA}
\author{David Neilsen}
\affiliation{Department of Physics and Astronomy, Brigham Young University, Provo, Utah 84602, USA}


\begin{abstract}
We study the merger of binary neutron stars with different
mass ratios adopting three different realistic, microphysical
nuclear equations of state, as well as incorporating neutrino cooling effects. In particular, 
we concentrate on the influence of the equation of state on the gravitational wave signature and also on its role, in combination
with neutrino cooling, in determining the properties of the resulting
hypermassive neutron star, of the neutrinos produced, and
of the ejected material.
The ejecta we find are consistent with other recent studies 
that find that small mass ratios produce more ejecta than equal mass cases (up to some limit) 
and this ejecta is more neutron rich.
This trend indicates the importance with future kilonovae observations of measuring the individual masses of an associated binary neutron star system,
presumably from concurrent gravitational wave observations, in order to be able to extract information
about the nuclear equation of state.
\end{abstract}

\maketitle

\tableofcontents

\section{Introduction}\label{introduction}
The spectacular detection of gravitational waves in event GW150914~\cite{Abbott:2016blz} marks
the start of the gravitational wave astronomy era. 
While this {\em first direct detection}
corresponds to a signal produced by the merger of a binary black hole system, the existence of gravitational
waves and our ability to detect them have been 
dramatically demonstrated. 
This observation will surely intensify efforts to identify potential signals in the various
bands (gravitational wave, electromagnetic wave, and even neutrinos for sufficiently close events). An early example of
this tantalizing  possibility is the possible---and puzzling---connection of  GW150914 with an 
observation by Fermi Gamma-ray Burst Monitor (GBM) ~\cite{Connaughton:2016umz} (see however~\cite{Evans:2016mta,Fermi-LAT:2016qqr}). 
This new era of multimessenger astronomy
will not only help test gravity and unravel spectacular astrophysics
events but, importantly, enhance the detection possibilities themselves. The flurry of activity
prompted by GW150914 serves as a clear example of what this era will bring.  
For instance, just to name a few examples, this single GW observation has spurred efforts to: test the basic tenants 
of General Relativity (e.g.~\cite{TheLIGOScientific:2016src,Blas:2016qmn,Ellis:2016rrr}),
infer astrophysical consequences on binary black hole population (e.g.~\cite{TheLIGOScientific:2016htt,Abbott:2016nhf}),
discuss possible mechanisms to explain the rather large stellar masses extracted from the signal (e.g.~\cite{Mandel:2015qlu,Marchant:2016wow}),
and re-examine how to possibly channel electromagnetic counterparts in stellar mass binary black holes (e.g.~\cite{Loeb:2016fzn,Perna:2016jqh}).

Among compact binary systems, binary neutron star systems present very exciting  
prospects both in terms of providing strong signals and for the underlying physics able to be explored.
For example,  tidal effects during
the late orbiting stages are imprinted on the gravitational wave signal and the interaction of the stars with
their surrounding plasma can power pulsar-like 
electromagnetic signals from radio to gamma ray 
wavelengths~\cite{Palenzuela:2013hu,Palenzuela:2013kra,Ponce:2014sza}. The merger of such a system produces a   
differentially rotating, hot massive neutron star~(MNS) with a strong neutrino luminosity. Depending
on the masses involved and the EOS describing the stars, this massive remnant may collapse to a black hole, allowing for
signals to reveal the ``birth cry'' of a black hole~\cite{Lehner:2011aa}. Such a black hole could lead to hyper-accretion and 
a short gamma-ray burst~(sGRB) or related phenomena. The merger itself could produce
significant outflow such that the ejected material might power a kilonova. A first reported of
kilonova observation has been presented in~\cite{2013Natur.500..547T,Berger:2013wna}. Furthermore,
binary neutron star systems are important to test for alternative theories
of gravity describing dynamics that depart significantly from the expected one in General Relativity, even 
when binary black holes behave precisely as in GR~\cite{Barausse:2012da,Palenzuela:2013hsa,Shibata:2013pra}.
Interestingly, such departures might be difficult to extract solely via gravitational waves~\cite{Sampson:2014qqa} and
complementary electromagnetic information will be important for such a task~\cite{Ponce:2014hha}.

The dynamics and identification of potential signals of such a complicated system are complex, generally requiring
numerical simulation of dynamical gravity, magnetized plasma, matter described with a realistic EoS, and finally
the ability to track the dominant nuclear reactions involving neutrinos.
Here we present the results of our latest studies of the merger of systems of two, unequal mass, neutron stars.

Our results suggest that unequal mass binaries, relative to the equal mass case, produce
more ejecta that is cool and neutron rich because the ejecta has been tidally ripped from the
constituent stars. That this material is neutron rich provides a likely environment for
r-process nucleosynthesis and may produce an afterglow consistent with a kilonova.
Additionally,
a radio counterpart might be produced as this material interacts with the interstellar medium surrounding
the binary. We provide estimates of the strength and time scales for such counterparts.
Contrary to the equal mass case, we note that essentially all unequal mass cases studied produce enough ejecta that
is sufficiently neutron rich for a related kilonova to peak in the infrared. 
This indicates that individual
mass estimates through gravitational wave information may be crucial to extract EoS information
using kilonovae observations from generic binary neutron star mergers 
We illustrate difficulties
in obtaining EoS information just from gravitational waveforms based on a small number of events unless post-merger detection and analysis is
within reach. In such a case, if prompt collapse to black hole does not take place, extraction of the dominant gravitational
 wave mode produced by the remnant can provide
a clean connection with the neutron star equation of state, and we present a simple fit
connecting the (possibly) observable frequency with the star's characteristics. In addition, we estimate neutrino luminosities and their dependence on EoS and mass ratio and discuss possible features that can potentially provide complementary
information on the EoS. 
We summarize details in section~II, present results in section~III, and discuss these results in section~IV.

\section{Numerical Implementation}
\label{sec:implementation}

The evolution equations for the spacetime and the fluid are already described in our previous paper\cite{2015PhRvD..92d4045P},
while full details of our implementation are described in~\cite{Neilsen:2014hha}. Unless
otherwise specified, we adopt geometrized units  where $G=c=M_\odot = 1$, except for
some particular results which are reported more naturally in physical cgs units.

Our stars consist of a fluid described by the stress energy tensor
\begin{equation}
T_{ab} = h u_a u_b + P g_{ab}  \, ,
\end{equation}
where $h$ is the fluid's {\it total} enthalpy $h \equiv  \rho (1+\epsilon) + P$,
and $\{\rho,\epsilon,Y_e,u^a,P\}$ are the rest mass energy density,
specific internal energy, electron fraction (describing the relative abundance of electrons compared to the total number of baryons), four-velocity, and pressure of the fluid,
respectively.  

To track the composition of the fluid and the emission of neutrinos we employ a leakage scheme.
In particular, the equations of motion consist of the following conservation laws
\begin{align}
\nabla_a T^a_b &= {\cal G}_b\, , \label{eq:DT}\\
\nabla^a (T_{ab} n^b) &= 0\, , \label{eq:DTN} \\
\nabla_a (Y_e \rho u^a) &= \rho R_Y\, , \label{eq:divYe}\,.
\end{align}
The sources ${\cal
  G}_a$~($\equiv -\nabla_c {\cal R}^c_{a}$) and $R_Y$ correspond to the radiation
four-force density and lepton sources and are computed within the
leakage scheme.
These equations are conservation laws for the stress-energy tensor, 
matter, and lepton number, respectively.
 Notice that, in the absence of lepton source terms ($R_Y=0$),
Eq.~(\ref{eq:divYe}) provides a conservation law for leptons, and is similar to
the familiar baryon conservation law, i.e., $Y_e$ is a mass scalar.

\section{Results}
\label{sec:numericalresults}

We extend our
previous work~\cite{2015PhRvD..92d4045P} and study the merger of {\em unequal
mass neutron star binaries}. In particular,
we concentrate on systems with parameters consistent with current observations (see e.g.~\cite{doi:10.1146/annurev-nucl-102711-095018}). 
Therefore, we will consider binaries with the same total gravitational mass $M=2.70 M_{\odot}$ and with different mass ratios $q \equiv M_1/M_2$ between $1$ (the equal mass case) and $0.76$

We continue with the same three distinct microphysical EoS
 (SFHo~\cite{2013ApJ...774...17S}, DD2~\cite{2012ApJ...748...70H}
and NL3~\cite{2012ApJ...748...70H}) used in our previous paper~\cite{2015PhRvD..92d4045P}. 
These three EoS were chosen to span a range of stiffnesses and they produce cold neutron 
stars with radii ranging from $\approx 12$~km  with the SFHo soft EoS to $\approx 15$~km
 with the NL3 stiff EoS
(for the same mass of $1.35 M_{\odot}$). 
All three EoS are consistent with  neutron stars having a  mass of at least 2\,$M_\odot$, and they are 
therefore consistent with observations of NS masses~\cite{Demorest:2010bx,Antoniadis:2013pzd}.

We focus here on the late stages of the coalescence, 
comprising the last $4$--$6$ orbits (depending on EoS), of
irrotational binary neutron star systems. The physical parameters of the binaries 
and our grid setup are summarized in Table~\ref{table:equal_mass}.

\begin{table*}[t]\centering
\begin{ruledtabular}
\begin{tabular}{lllllllllllll}
EoS & q & $\nu$ 
 &  $m_{b}^{(1)}, m_{g}^{(1)}$ &  $m_{b}^{(2)}, m_{g}^{(2)}$  
  & $R^{(1)}$  & $R^{(2)}$ 
  & $C^{(1)}$  & $C^{(2)}$  
 & $J_{0}^{\rm ADM} $ 
 & $\Omega_0$ & $f_0^{\rm GW}$  
 & $M_{\rm eject}$ \\ 

 &  &  & ~~[$M_{\odot}$] &  ~~[$M_{\odot}$]  
 & [km] & [km] &  &   
 & [G $M^2_{\odot}/c$] 
 & [rad/s] & [Hz]  
 & [$10^{-3} M_{\odot}$] \\

\hline
 NL3  & 1.0  & 0.250 &  1.47, 1.36  & 1.47, 1.36 & 14.80 & 14.80 & 0.136 & 0.136 & 7.40 & 1778 & 566 & $0.015$
 \\ 
 NL3  & 0.85  & 0.248 & 1.34, 1.25 & 1.60, 1.47 & 14.75 & 14.8 & 0.125 & 0.147 & 7.35 & 1777 & 566 & $ 2.3$
  \\
 DD2  & 1.0  & 0.250 & 1.49, 1.36 & 1.49, 1.36 & 13.22 & 13.22 & 0.152 & 0.152 & 7.39 & 1776 & 565 & $0.43$
  \\ 
 DD2  & 0.85  & 0.248 & 1.36, 1.29 & 1.62, 1.47 & 13.20 & 13.25 & 0.144 & 0.164 &  7.34 & 1775 & 565 & $0.42$
  \\ 
 DD2  & 0.76 & 0.245 & 1.27, 1.18 & 1.71, 1.54 & 13.16 & 13.25 & 0.132 & 0.172  & 7.26 & 1775 & 565 & $1.3$
  \\ 
 SFHo & 1.0  & 0.250 & 1.50, 1.36 & 1.50, 1.36  & 11.90 & 11.90 & 0.169 & 0.169 & 7.38 & 1775 & 565 & $3.4$
 \\ 
 SFHo  & 0.85 & 0.248 & 1.37, 1.25 & 1.63, 1.47 & 11.95 & 11.85 & 0.154 & 0.183 & 7.31 & 1773 & 564 & $2.2$
  \\ 

\end{tabular}
\end{ruledtabular}
\caption{Summary of the binary neutron star systems considered in
this work. The initial data were computed using the {\sc Bin star}
solver from the {\sc Lorene} package~\cite{lorene}, with the
assumption that the stars have an initial constant temperature of
$T=0.02$~MeV and are in beta-equilibrium. All the binaries have a
total mass $M_{0}^{\rm ADM} = 2.7 M_{\odot}$ and start from an
initial separation of $45$~km. The outer boundary is located at
$750$~km and the highest resolution covering both stars is $\Delta
x_{\rm min}=230 m$. The table displays the mass ratio of the binary
$q \equiv M_1/M_2$ and $\nu=M_1 M_2/M^2 = q/(q+1)^2$, the baryon (gravitational) mass
of each star $m_b^{(i)}$ ($m_g^{(i)}$), its circumferential radius $R^{(i)}$ and
its compactness $C^{(i)}$ (i.e., when the stars are at infinite
separation), the angular momentum $J_{0}^{\rm ADM}$, the initial
orbital angular frequency $\Omega_0$, the initial GW frequency
$f_0^{\rm GW} \equiv \Omega_0/\pi$ of the system, and finally the
total mass ejected during the merger process, $M_{\rm eject}$.
}
\label{table:equal_mass}
\end{table*}

\subsection{Gravitational Waves}
Before describing the characteristics of the GW signals, consider first the 2-body, Post-Newtonian approximation
for a compact binary system (as can be found in, for example, Ref.~\cite{2014grav.book.....P}).
This approximation describes the interaction via an expansion with respect to the parameter
$x \equiv (M \omega)^{2/3}$ where $M$ is the total mass and $\omega$ is the orbital frequency. 
At the lowest order, $x=M/R$ with $R$ the separation of the binary.

The amplitude of the gravitational waves produced, $h$ (and consequently the energy radiated and angular momentum loss),
from the system are, to leading  order, given by the scaling
$h\propto {\cal M}^{5/3} \omega^{2/3}/r$.
Here the chirp-mass is defined as ${\cal M} \equiv (m_1 m_2)^{3/5}/(m_1+m_2)^{1/5}$ and $r$ is the distance
to the source. Higher order contributions containing both conservative and dissipative pieces, depending
on the constituent masses, spin, radiation-reaction, and internal structure, 
become increasingly important as the frequency (separation) increases (decreases) (see e.g.~\cite{2014grav.book.....P}). 

Particularly relevant for
our current discussion are tidal effects and the effect of mass ratio.
For the former, given a mass ratio~\footnote{We adopt the convention $m_2 \ge m_1$ so that $q\le 1$.} $q\equiv m_1/m_2$, the chirp mass 
is ${\cal M} = M q^{3/5}/(1+q)^{6/5}$ [which is also $= 2^{-6/5} M \left( 1  - 3\delta^2/20  + {\cal O}\left(\delta^3 \right)\right) $ in terms of 
$\delta \equiv 1 - q$] .  
Thus the dominant contribution to the gravitational wave amplitude 
is already sensitive to the mass ratio. 

However, for realistic binary neutron star systems, the parameter $q$ is limited
to a small range and $3 \delta^2/20 \simeq \{0 ,3 \times 10^{-3},9 \times 10^{-3}\}$ for $q=\{1,0.85,0.76\}$
thus mass ratio differences in BNS systems are small. 

Tidal effects, on the other hand, can become quite significant at close separations, appearing with
a high power of the inverse of the stellar compaction.  In particular, even though tidal effects enter
the PN approximation at tenth order in $x$, this sub-leading effect can still be apparent
at sufficiently high frequencies since $C_i \in (0.05 - 0.2)$ for neutron stars.
That is, the leading contribution to the tidal force on a single star is given by
\begin{equation}
F_T \propto  k_2^{(i)} C_i^{-5} x^{10} 
\end{equation}
where $k_2^{(i)}$ is the quadrupolar tidal coupling Love number constant and $C_i=M_i/R_i$ (for the $i$th star)~\cite{Vines:2011ud,Mora:2003wt}. 
The dependence of tidal effects on the binary's dynamics
can be expressed via $\kappa_2^T$ for the binary in terms of the individual $ k_2^{(i)}$, defined as~\cite{2010PhRvD..81h4016D,2015PhRvD..91l4041D}
\begin{equation}
  \kappa_{2}^{T} = 2 \left[ \frac{q^4}{(1+q)^5} \frac{k_2^{(1)}}{C_1^5} + 
  \frac{q}{(1+q)^5} \frac{k_2^{(2)}}{C_2^5}  \right].
\end{equation}
For neutron stars, the Love number constant typically $\in (10^{-2},0.5)$~\cite{Hinderer:2007mb} 
and their compaction ratios $C \simeq 0.1 - 0.2$. Thus for sufficiently high frequencies (close separations) tidal
effects can become important as $\kappa_{2}^{T} x^{10} \simeq {\cal O}(1)$. Notice that  for a fixed given total
mass, stiffer EoS (i.e., with larger radii and smaller compactness) 
lead to larger $\kappa_2^{T}$ and therefore stronger tidal effects. 
Similarly,  unequal mass cases also lead to slightly larger $\kappa_2^{T}$ thus enhancing
tidal effects.
For low frequencies on the other hand,  $\kappa_{2}^{T} x^{10} \ll {\cal O}(1)$, and only the mass ratio can introduce departures when comparing
binaries with the same total mass. However, these departures are necessarily small for BNS systems. 

\begin{figure}[h]
\centering
\includegraphics[width=8.5cm,angle=0]{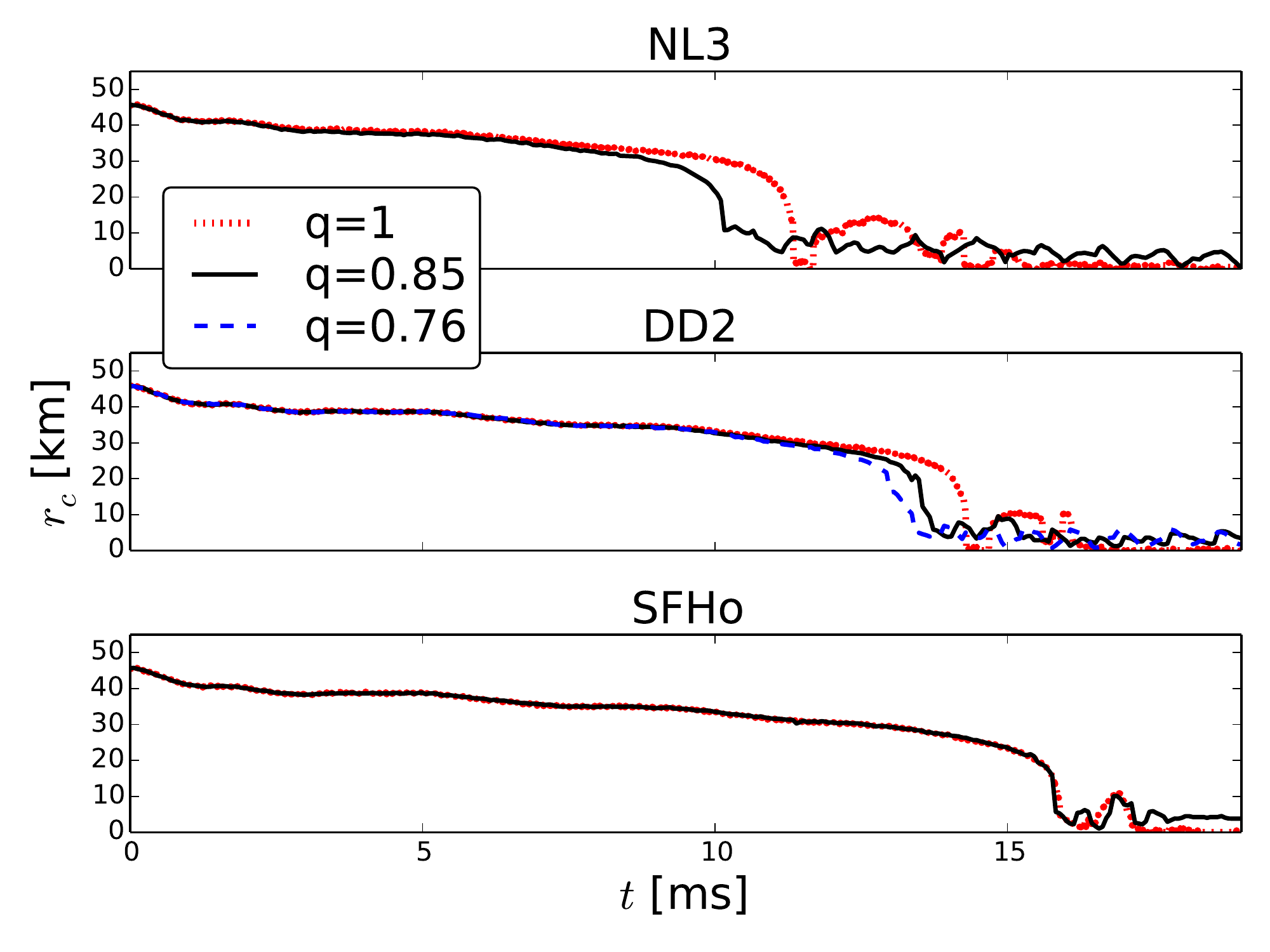}
\caption{ The coordinate separation between the centers of the stars as a
function of time for different mass ratios and EoS. 
Overall, unequal mass binaries merge more quickly than equal mass binaries,
and the time to merger increases for softer equations of state with smaller 
stellar radii.
The coordinate distance is
approximated by the positions of the maximum densities.
}
\label{fig:eos_separation}
\end{figure}

\begin{figure}[h]
\centering
\includegraphics[width=8.5cm,angle=0]{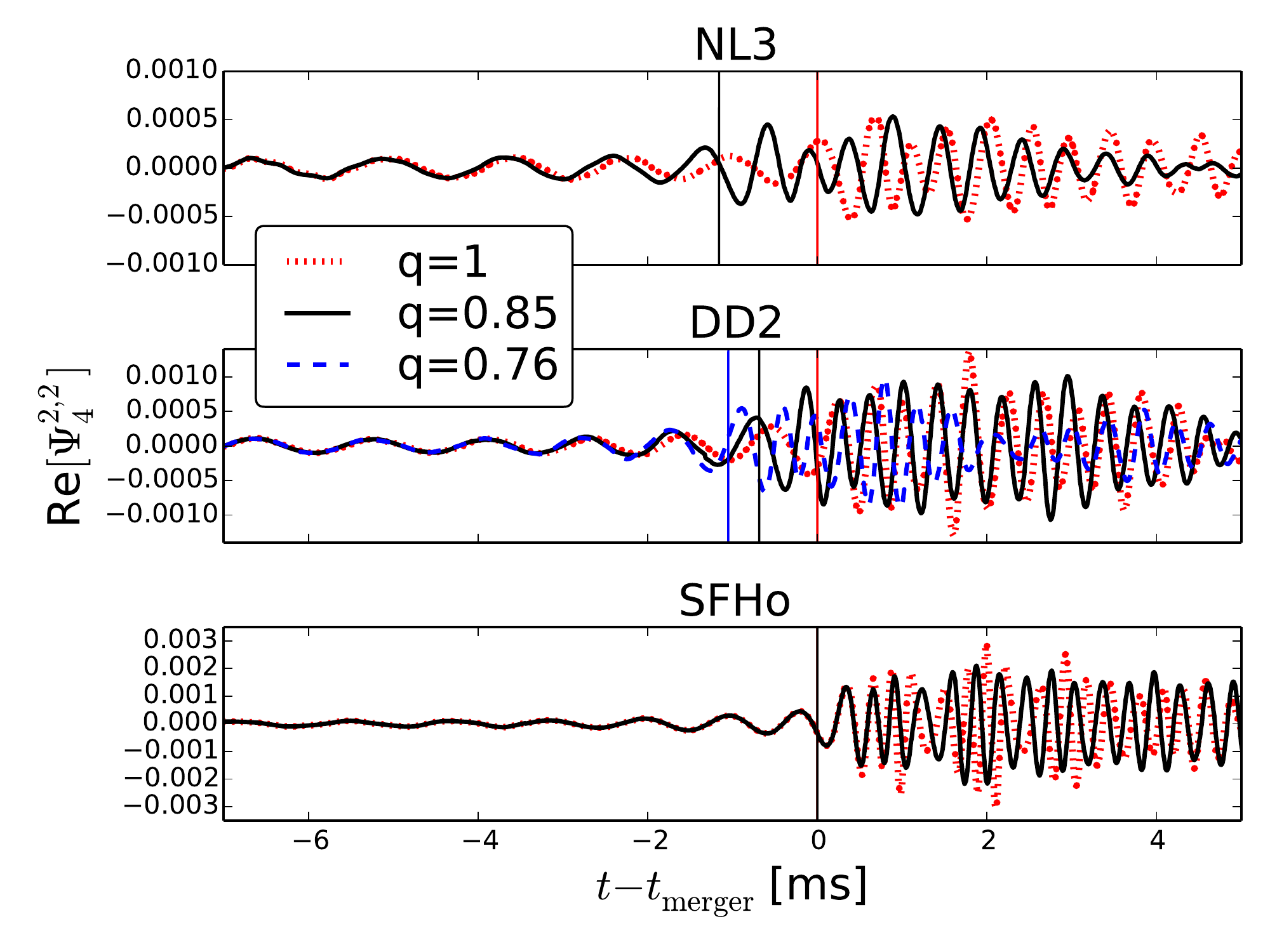}
\caption{A comparison of the gravitational wave signals near merger using the
primary mode of $\Psi_4$. The signals have been shifted in time such that
$t=0$ corresponds to the moment of contact for $q=1$, with respect to each EoS, 
and vertical bars indicate the moment of contact for the other cases [the time labels
are the same in all panels].
For NL3 and to a lesser extent for DD2, clear differences in the waveforms 
arise prior to the stars coming into contact. 
In contrast, the SFHo waveforms show essentially no differences before merger. 
After the merger however, there are significant differences between all of the 
waveforms, showing that they carry information about the post-merger star,
e.g., internal structure, radius, compactness, etc., 
that is sensitive to the equation of state.
}
\label{fig:eos_waveform}
\end{figure}

The consequences of the above observations are clearly illustrated in Figs.~\ref{fig:eos_separation} and~\ref{fig:eos_waveform}. Fig.~\ref{fig:eos_separation} shows 
the (coordinate) separation versus time for all EoS and mass ratios studied while Fig.~\ref{fig:eos_waveform} displays the gravitational
wave signals as computed by the Newman-Penrose scalar $\Psi_4$. 

The times in Fig.~\ref{fig:eos_waveform} have been relabeled in terms of a merger time.
For each EoS, all mass ratios use the same time. In particular, the merger time is the time at which,
for the $q=1$ case,
the two locations of density maxima (the centers of the two stars) decreases below a threshold
distance (generally the sum of the two stellar radii), corresponding to the stars making contact.
Because the signals in each panel of Fig.~\ref{fig:eos_waveform} share the same time, we have included
a vertical
line to denote the time when such contact occurs for the unequal mass cases.  
As can be clearly appreciated in both figures,  the early behavior of all mass ratios within a given
EoS ($<7$\,ms) agree.

However, as the orbit tightens tidal deformations increase and affect the dynamics of the system. Their
effects manifest first for binaries described by the NL3, followed by the DD2, and finally the SFHo equations of state.
As expected from our discussion above, this succession is tied to the decreasing radii characterizing the stars of 
each binary. For instance, Fig.~\ref{fig:eos_separation} indicates that a more rapid plunge takes
place for {stars with} larger radii, and, interestingly, within the same EoS the merger takes place earlier as 
the mass ratio is decreased. This behavior is apparent in the gravitational wave signatures
in which binaries with large stellar radii demonstrate differences earlier than those binaries with smaller radii.
Furthermore, comparison within the same EoS reveals that, as the mass ratio decreases,
an earlier departure from a simpler (chirping) sinusoidal wave arises as tidal effects become more important.
This pre-merger behavior is especially clear for the NL3 case but is essentially negligible for the softer EoS considered.

Additionally $q\ne 1$ cases also display a stronger $l=2,m=1$ component in the gravitational waves during the late pre-merger
and afterwards than the $q=1$ case. However, such modes are still at least a couple of orders of magnitude weaker than the main $l=2,m=2$ signal displayed in Fig.~\ref{fig:eos_waveform}.
Nevertheless, they can strongly seed a possible $m=1$ instability after the merger whose associated frequency would
be lower than the one corresponding to $m=2$ modes~\cite{East:2015vix,Chawla:2010sw}.

The post-merger behavior
is governed by the properties of the resulting hot, differentially rotating MNS (or hypermassive NS) which 
loses energy and angular momentum via emission of neutrinos and gravitational waves. Depending
on the EoS and the total mass, the remnant MNS may eventually collapse to a black hole or continue as a rotating
star. In cases that avoid prompt collapse to a black hole~\footnote{Remnants with  total mass not significantly
larger than the maximum allowed mass for a rotating neutron star for a given EOS can avoid prompt collapse, e.g~\cite{Hotokezaka:2011dh}.},
gravitational waves in this stage are characterized by a few dominant modes with frequencies intimately tied to the EoS.  
These dominant modes arise are determined by the natural oscillation frequencies of the star, by the details of the merger process itself, and by the possible
development of an $m=1$ instability if a rather long-lived MNS is possible. Naturally, these modes
are a potentially rich  source of information about the system (see for instance~\cite{lrr-1999-2} and references within).

\begin{figure}[h]
\centering
\includegraphics[width=8.2cm,angle=0]{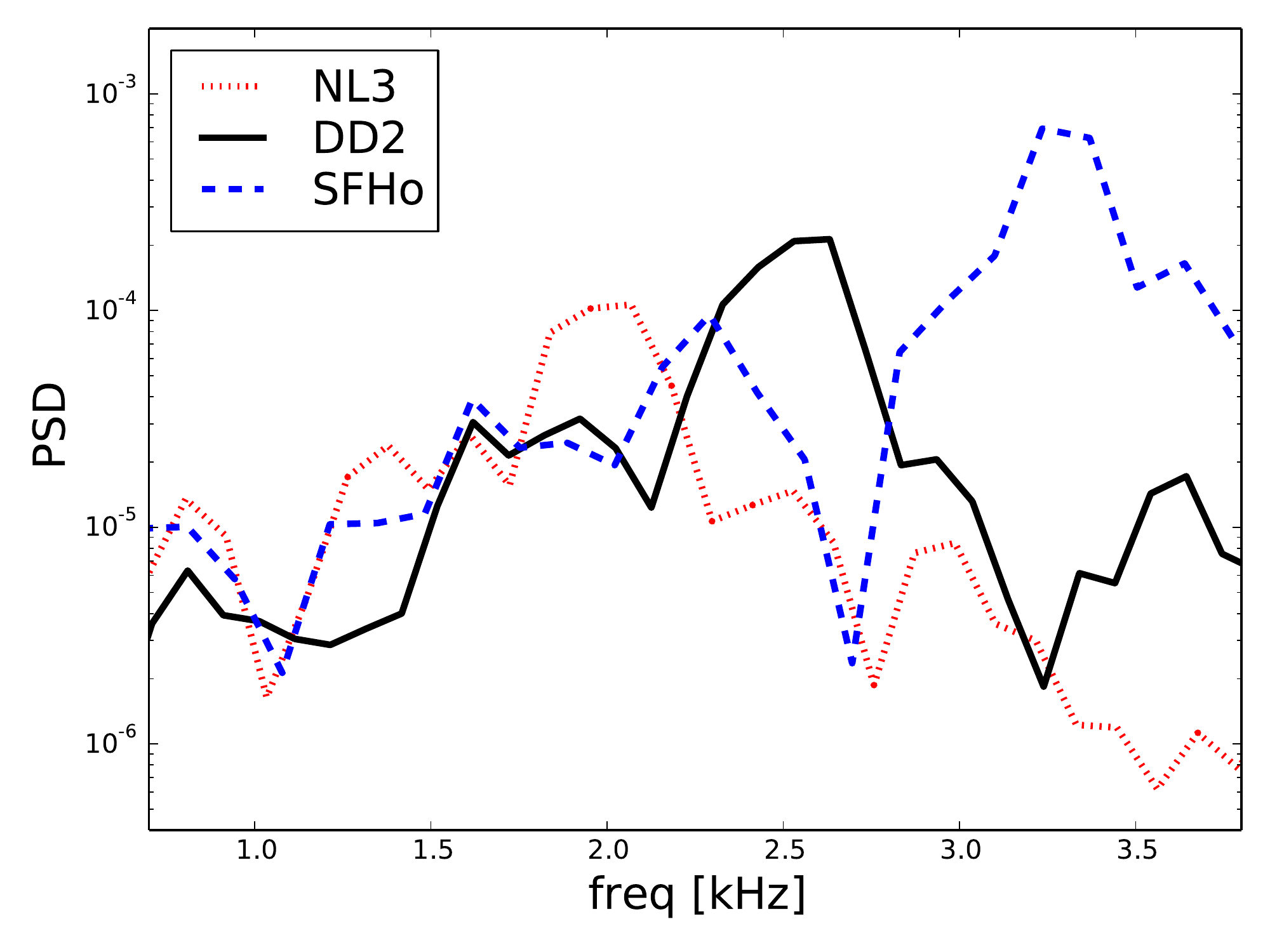}
\includegraphics[width=8.2cm,angle=0]{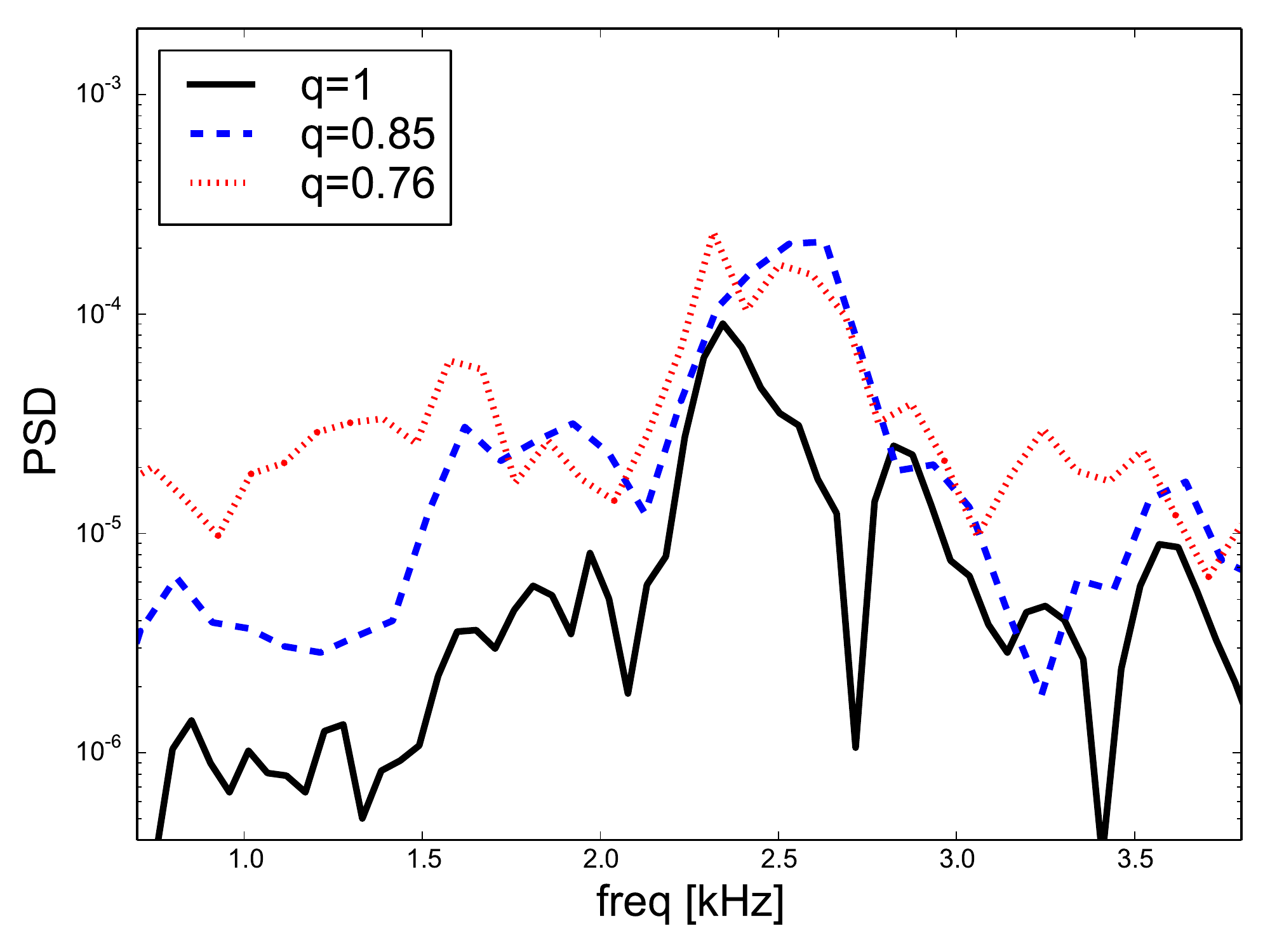}
\caption{ Power spectral densities~(PSD) of the gravitational waveforms after
the BNS merger for the three different equations of state. 
The peaks in the Fourier transform of the waveforms represent the dominant
oscillation modes of the MNS, $f_i$, presented in Table~\ref{table:modes_HMNS}.
(Top:) The PSDs for the $q=0.85$ cases with the three different EoS.
(Bottom:) The PSDs for three values of $q$ for just the DD2 EoS.
} 
\label{fig:eoswaveform_postmerger}
\end{figure}

We scrutinize these modes more closely by computing the Fourier transform of the
primary $\Psi_4$ gravitational wave mode ($l=2=m$) in a time window of duration roughly $10\,$ms after the merger.
These spectra, shown in Fig.~\ref{fig:eoswaveform_postmerger}, demonstrate several characteristic peaks, and these peak frequencies 
are presented in  Table~\ref{table:modes_HMNS}.  The most dominant of these modes, using the terminology of Ref.~\cite{2015arXiv150203176B}, $f_{\rm peak}$, is associated with the rotational
behavior and quadrupolar structure of the newly formed MNS/HMNS. Secondary modes, at lower 
frequencies, can be associated with different 
mechanisms~\cite{2015arXiv150203176B,2015PhRvD..91f4027K,2015PhRvD..91f4001T,East:2015vix}.
However as noted in~\cite{2015arXiv150203176B}, in low total mass binaries~\cite{Foucart:2015gaa} and for the
case of a stiff EoS as in Ref.~\cite{2015arXiv150908804D},
(the majority of) these secondary modes degrade quickly with time in just a few milliseconds and are unlikely
to build enough SNR for detection. The primary mode on the other hand,
can last for long times and provide the best opportunity for detection. Consequently it has been
the focus of increased scrutiny.

Recently, a fit for these modes has been suggested for the case of equal mass binaries~\cite{2015arXiv150203176B}, 
and we include in Table~\ref{table:modes_HMNS} the frequencies predicted from this fit together with  our measured values.  
As discussed in our previous work~\cite{2015PhRvD..92d4045P},
our results for $q=1$ are within $10\%$ of the predicted frequencies.

A similar fit for unequal mass binaries described by tabulated EoS is so far unavailable because, contrary to the equal mass case, studies with $q\ne1$ are still in their infancy. The majority
of currently available results for unequal mass binaries employ piece-wise polytropes (chosen to be reasonable approximations to tabulated, realistic EoS.). A comparison of results obtained with
the so-called MS1, H4, SLy cases, studied in~\cite{2015PhRvL.115i1101B}, which have similar
mass ratios and masses to those used here for the NL3, DD2, SFHo EoS, reveals
that the peak frequencies agree to better than $5\%$. 
As well, Ref.~\cite{2015arXiv150908804D} found that, for the polytrope dubbed SLyPP
(which yields a NS similar to the one described by the NL3 EoS considered here),
the resulting peak frequency, $f_{\rm peak}$, has a small dependence on the mass ratio, with
a slightly higher frequency for the equal mass case. This finding is also consistent with our
results. Last, relatively small differences in the peak frequency with respect to mass ratio
have also been reported in other recent 
works~\cite{2015arXiv150203176B,2015arXiv150707100D,2015PhRvD..91l4041D}.
As can be appreciated in Table~\ref{table:modes_HMNS}, we 
also find that the peak frequencies for $q\ne 1$ lie
 within $\approx 10\%$ of the corresponding $q=1$ case and are typically lower ---except for $q=0.85$ DD2.

Beyond comparing with particular fits, 
it is instructive to take a step back and consider estimating the peak frequency based on simple arguments which
can shed light on the underlying physics of the system. A simple approach is to consider the contact
frequencies, $f_c$,  of the binary, based on the radii of the two stars.  This approach can
only provide a lower bound but it may potentially lead to improved estimates in the future.
The contact frequency~\footnote{As a certain amount of compression results
in the collision, the true gravitational frequency is naturally higher.} is given in terms of the binary parameters by  (in $M^{-1}$ units) 
\begin{equation}
  f_{\rm c} = \frac{1}{\pi M_g} \left( \frac{m_g^{(1)}}{M_g C_1} + \frac{m_g^{(2)}}{M_g C_2}  \right)^{-3/2} 
\label{fcontact}
\end{equation}
where
$M_g=m_g^{(1)}+m_g^{(2)}$, (see Eq.~(36) of Ref.~\cite{Damour:2012yf}).
We include with our measured frequencies the estimated contact frequencies in
Table \ref{table:modes_HMNS}. 
While they are clearly lower than the observed frequencies as anticipated, they do provide a consistent trend. 

\begin{figure}[h]
\centering
\includegraphics[width=8.2cm,angle=0,clip]{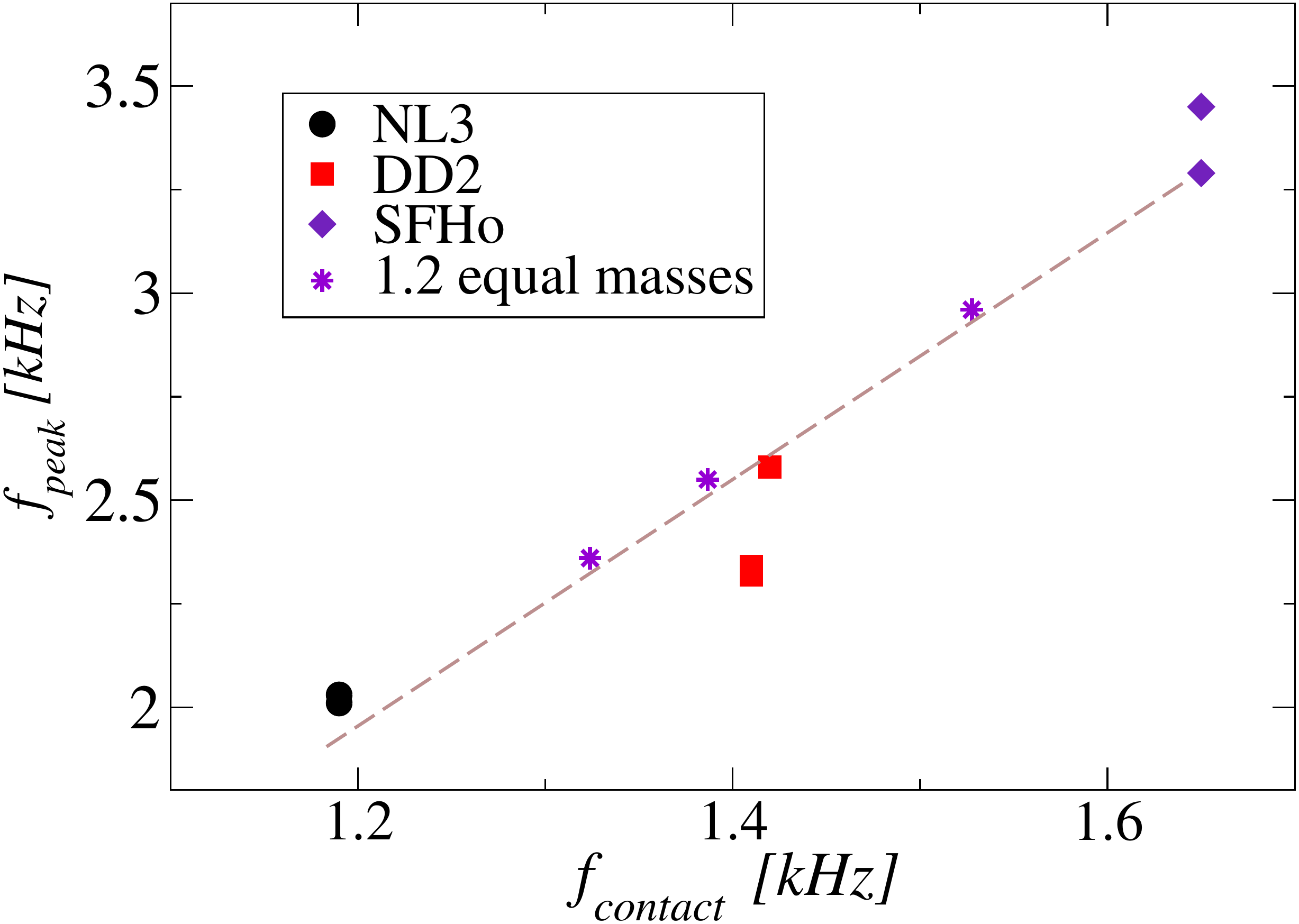}
\caption{After-merger peak-frequency computed via simulations
versus the estimated contact frequency for several BNS mergers. 
In addition to the quantities calculated in this paper, we include results 
for an equal-mass binary neutron 
star system with a smaller total mass from Ref.~\cite{Foucart:2015gaa}, which
used the DD2, LS220, and SFHo equations of state.}
\label{fig:fpeakvsfestimate}
\end{figure}

This trend
is more clearly seen in Fig.\ref{fig:fpeakvsfestimate} which illustrates the measured
peak frequency versus the estimated contact frequency for all cases studied here (which corresponds
to total mass $M=2.7M_{\odot}$) together with data from Ref.~\cite{2015arXiv150203176B} (for
equal mass binaries with $M=2.4M_{\odot}$). The observed trend can be expressed quantitatively
through a linear regression fit to our data
\begin{equation}
f_{\rm peak} [\mathrm{kHz}] = -1.61 + 2.96 f_{\rm c}\left[ \frac{2.7 M_\odot}{M} \right] [\mathrm{kHz}]. \label{fitestimate}
\end{equation}
This correlation is characterized by a correlation coefficient $=0.96$ and has been
obtained with total mass $M=2.7M_{\odot}$. For systems with  a different total mass, Eq.~(\ref{fitestimate})
effectively rescales the contact frequency.

For instance, Ref.~\cite{Foucart:2015gaa} presented peak frequencies from evolutions of equal mass binaries with total mass
$M=2.4M_{\odot}$ 
concentrating on three EoS: the same DD2 and SFHo EoS studied here along with
LS220~\cite{lattimer_swesty} which yields radii in between the other two for the same mass. 
The corresponding peak frequencies
computed for DD2, LS220, and SFHo EoS are $2.35\,\mathrm{kHz}$, $2.56\,\mathrm{kHz}$, and $2.96\,\mathrm{kHz}$, respectively. 
Expression~(\ref{fitestimate}) 
estimates the
peak frequencies to be $f_{\rm peak}=2.32\,\mathrm{kHz}$, $2.49\,\mathrm{kHz}$, and $2.91\,\mathrm{kHz}$, respectively, for these cases. 
The agreement is quite good. It would be interesting to explore if this simple approach
works as well in the case of spinning binaries (e.g.~\cite{Tacik:2015tja,East:2015vix}) 
and to contrast it to alternative
fits depending explicitly on tidal deformability parameters~\cite{Bernuzzi:2014owa}.
%

%

\begin{table}[t]\centering

\begin{ruledtabular}
\begin{tabular}{llllllllll}
EoS & q &  $f_{1}$ &  $f_{2}$  & $f_{3}$ & $f_{4}$  
&  $f_{\rm peak}$  & $f_{\rm spiral}$ & $f_{2-0}$ & $f_{c}$ \\ 
\hline
NL3  & 1.0 & 2.03 & 1.54  & 0.83  & --   &  2.2  & 1.6  & 1.2 & 1.19 \\
NL3  & 0.85 & 2.01 & 1.61  & 1.37  & 0.8  & --  & --  & -- & 1.19 \\ 
DD2  & 1.0 & 2.34 & 1.97  & 1.82  & 1.62 & 2.6  & 1.9  & 1.5 & 1.41 \\
DD2  & 0.85 & 2.58 & 1.92  & 1.62  & -- & --  & --  & -- & 1.42 \\ 
DD2  & 0.76 & 2.32 & 1.86  & 1.62  & --  & --  & --  & -- & 1.41  \\ 
SFHo & 1.0 & 3.45 & 2.59  &  2.20 & 1.62  & 3.2 & 2.4  & 2.1 & 1.65 \\
SFHo & 0.85 & 3.29 & 2.29  &  1.61 & --  & -- & --  & -- & 1.65  \\
\end{tabular}
\end{ruledtabular}

\caption{Prominent oscillation frequencies (kHz) in the power spectral 
densities of the post-merger gravitational waveform compared with predicted
values. The frequencies $f_1$, $f_2$, $f_3$, and $f_4$. correspond to 
various peaks of the post-merger GW spectrum 
(see Fig.~\ref{fig:eoswaveform_postmerger}). 
$f_{\rm peak}$ and $f_{\rm spiral}$ are the predicted peak 
frequencies from Ref~\cite{2015arXiv150203176B}.
The correspondence between $f_1$ and $f_{\rm peak}$, 
$f_2$ and $f_{\rm spiral}$, and 
either $f_3$ or $f_4$ with $f_{2-0}$
suggests consistency with the model presented in~\cite{2015arXiv150203176B}
(which was tailored for the equal mass case, but reports errors $< 5\%$ for
mass ratios $q=0.92$). $f_c$ is the computed contact frequency~\ref{fcontact}.
}
\label{table:modes_HMNS}
\end{table}

Finally, beyond the specific dynamics and waveform characteristics,
it is instructive to  
explore the extent to which the obtained waveforms could be distinguished with
respect to different configurations of aLIGO. To do so, we first obtain a longer
wavetrain by employing at earlier times a PN (Taylor T4) expansion 
augmented to include leading order tidal effects as described in Refs.~\cite{Vines:2011ud,Hotokezaka:2013mm}.
We match this augmented PN signal at early times to
our numerical signal before merger to form a hybrid waveform. (Further improvements in the
early signal can be achieved, see e.g.~\cite{Bernuzzi:2014owa,Hinderer:2016eia}, though such improvements
would not affect the main conclusions drawn here.)
The matching is accomplished by choosing a cycle devoid of the initial numerical transients but still
well before merger. Within this cycle, a weighted average of the two waveforms is calculated
using  a tanh function that transitions from $0$ to $1$. 

\begin{figure}[h]
\centering
\includegraphics[width=9.2cm,angle=0,clip]{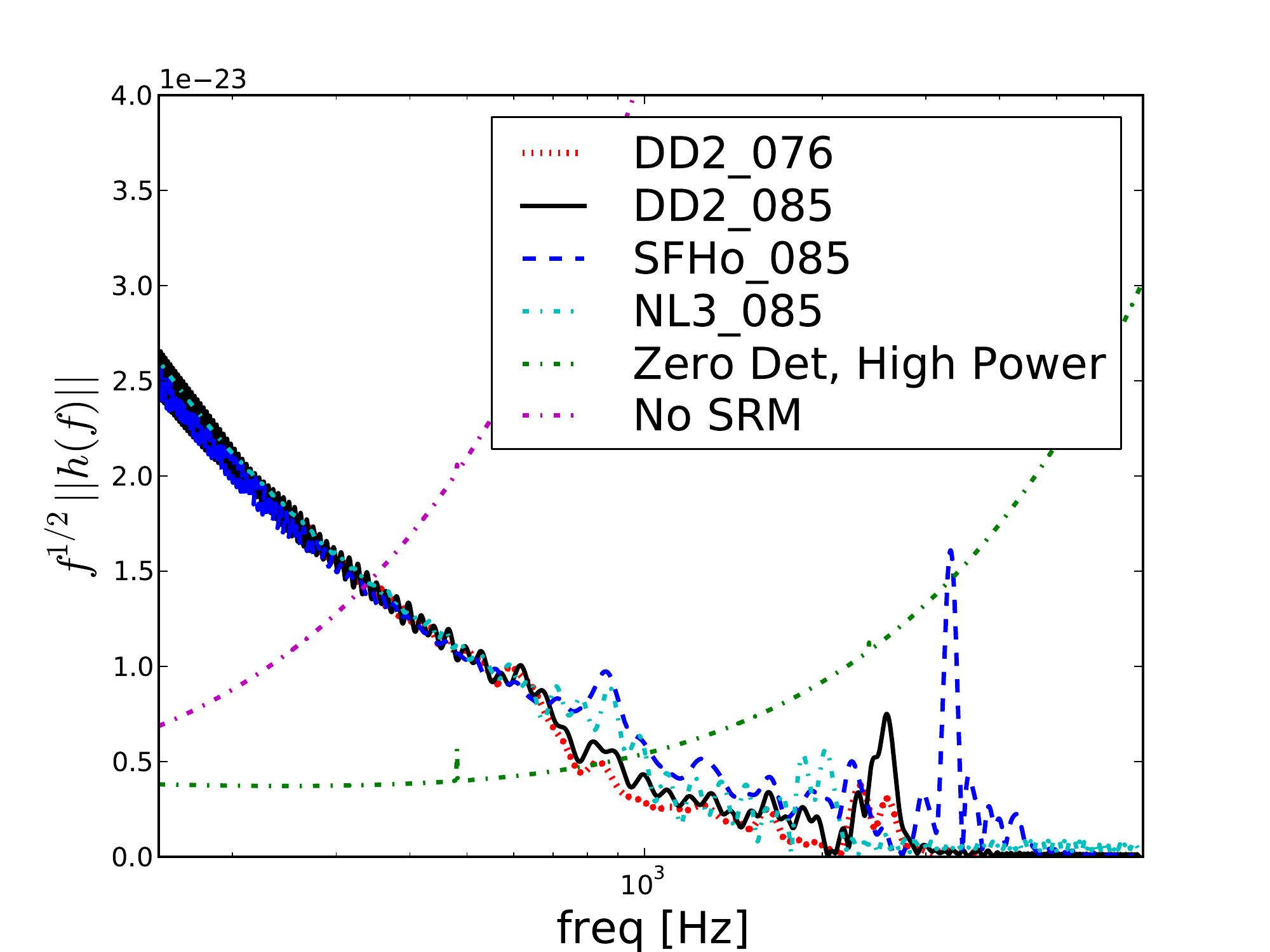}
\caption{The Fourier  spectrum  of  the GW signals
for the four binaries. The dotted and dashed-dotted monotonic
curves at high frequencies show
two fits to the noise power spectrum $\sqrt{S_h}$ of aLIGO (specifically the ``zero-detuned high-power'' and the
``No Signal Recycling Mirror''~(NSRM) cases~\cite{LIGOCURVE}.} 
\label{fig:noisecurves}
\end{figure}

Fig.~\ref{fig:noisecurves} illustrates the strain of the waveforms for the different cases
considered here with respect to two possible noise curves from aLIGO for binaries at 100\,Mpc. 
As can be appreciated in the figure, essentially no differences exist that could be discerned
with the {\em ``No SRM''} noise curve {\em with a single event at such a distance}. The more aggressive noise curve corresponding
to the {\em ``Zero Detuned, High Power''} configuration would, however, be sensitive to differences between
the stiffest and softest EoS considered. 

As a figure of merit, it is instructive to contrast the distinguishability of the obtained
waveforms. Since we start all our simulations at the same separation (orbital frequency)  we compute
the ``distinguishability''  between the different waveforms
(as described in e.g.~\cite{Read:2013zra}, see also~\cite{Owen:1995tm,Baumgarte:2006en,Read:2009yp})
using the {\em ``Zero Detuned, High Power''} and the {\em ``No SRM''}' noise curves 
in aLIGO~\cite{LIGOCURVE}. Table~\ref{tablematch} lists the values obtained for 
$\delta h \equiv \sqrt{(h_i - h_j, h_i - h_j)}$
(without allowing for frequency shifts and without normalizing our waveforms). 
Values above $1$ imply
such waveforms could be distinguished, indicating therefore that DD2 waveforms for $q=0.76$ could be distinguished
from all the others when using the more aggressive noise curve. The other cases are not discernible, even
when using the {\em Zero Detuned, High Power} noise curve.

The delicate nature of distinguishing EoS effects in binary neutron star systems has been thoroughly
discussed in Refs.~\cite{Lackey:2014fwa,Agathos:2015uaa}, which stressed the need for tens of
events to disentangling EoS differences if employing gravitational waves alone. Importantly,
as discussed in, e.g.~\cite{Bloom:2009vx,Branchesi:2011mi,Andersson:2013mrx,2012ApJ...746...48M,2015PhRvD..92d4045P,Radice:2016dwd}  and later in this work,
complementary electromagnetic information can help reveal the
EoS from the characteristics of  the radioactive decay of heavy elements produced by ejected
material.

\begin{table}[t]\centering

\begin{ruledtabular}
\begin{tabular}{l|lll}
EoS &  DD2, q085 &  SFHo, q085  & NL3, q085 \\ \hline
DD2, q076  &  1.70 (0.09) & 1.078 (0.05)  & 1.50 (0.07)  \\ 
DD2, q085  &  & 0.96 (0.05) & 0.77 (0.03)  \\ 
SFHo, q085  &  &   & 0.77 (0.03) \\ 
\end{tabular}
\end{ruledtabular}
\caption{The distinguishability between gravitational
waveforms for binary mergers with different equations of state for the 
zero-detuned high power (no signal-recycling mirror) noise curves for aLIGO.
}
\label{tablematch}
\end{table}

\subsection{Matter dynamics: Outflow and Ejecta}
\label{sec:outflows}
The merger of a binary with the total mass 
considered here produces a hot, rotating
neutron star that is stable for some time against gravitational collapse to a black hole
due to the rapid differential rotation and the thermal pressure. The merger
that produces such a star is quite violent and, as a by product, a significant
amount of material can be ejected from the binary. The characteristics of this
ejecta is important as it can produce an associated 
electromagnetic signal~\cite{Li:1998bw,Kulkarni:2005jw,2010MNRAS.406.2650M}
that can provide complementary information about the
binary (e.g.~\cite{Hotokezaka:2013kza,Piran:2014wpa,2015PhRvD..92d4045P,Foucart:2015gaa}).

The recent observation of an infrared transient associated associated with 
the SGRB 130603B~\cite{2013Natur.500..547T,Berger:2013wna} has demonstrated 
that such
a tantalizing prospect is within reach. Additionally as discussed in~\cite{Kasen:2014toa},
bound matter might also
induce observable electromagnetic signals through the properties of winds emitted by
the system~\cite{Kasen:2014toa,Fernandez:2014cna,Fernandez:2014bra}. 

We thus concentrate here on describing the properties of outflow material and its dependence
on both the nuclear EoS and the binary mass ratio. The latter is particularly relevant as
this material generally originates via tidal effects and is therefore especially sensitive to the mass ratio.
We analyze in detail the various binaries at $\sim$3\,ms after merger. A key distinction is made between matter
remaining bound to the system and ejecta, i.e. material having kinetic energy sufficient to
unbind it. Our analysis uses the common approach that material is considered
unbound if its energy (defined by assuming that the spacetime is stationary) is positive.

We note that an important issue when studying low-density material in simulations adopting 
conservative ``Eulerian'' schemes (as opposed to Lagrangian schemes) 
is that the results in low density regions could be
significantly affected by the artificial atmosphere employed specifically to handle such regions.
The simulations presented here benefit from an improvement to
our code's treatment of the atmosphere so that the floor value decreases rapidly with distance from the
center of coordinates (similar to that used in~\cite{2013PhRvD..87b4001H}). Comparisons 
of the new atmosphere treatment with the original, as well as studies with different atmosphere values, 
indicates that the properties being presented are nearly-independent of this improvement when 
the atmosphere density is sufficiently small $\approx 5 \times 10^{5}\,\mathrm{g}/\mathrm{cm}^3 \approx 10^{-10} \rho_\mathrm{max}$ 
and when estimates are calculated shortly after merger of $\leq 4-5$\,ms.

\begin{figure}[h]
\centering
\includegraphics[width=9.2cm,angle=0,clip]{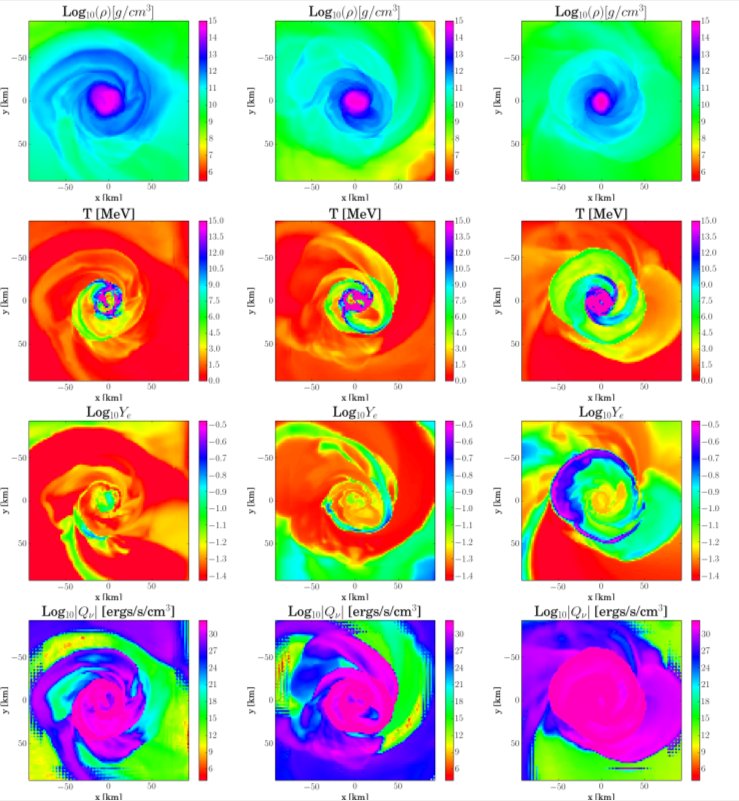}
\caption{Density, temperature, electron fraction, and neutrino
luminosity rate for the different EoS on the $z=0$ plane at roughly $t=3$~ms after the merger.
Shown are the NL3 EoS (left), DD2 EoS (middle), and SFHo (right)
for the mass ratio $q=0.85$. 
} 
\label{fig:eos_2dplots_zplane_q085}
\end{figure}

\begin{figure}[h]
\centering
\includegraphics[width=9.2cm,angle=0,clip]{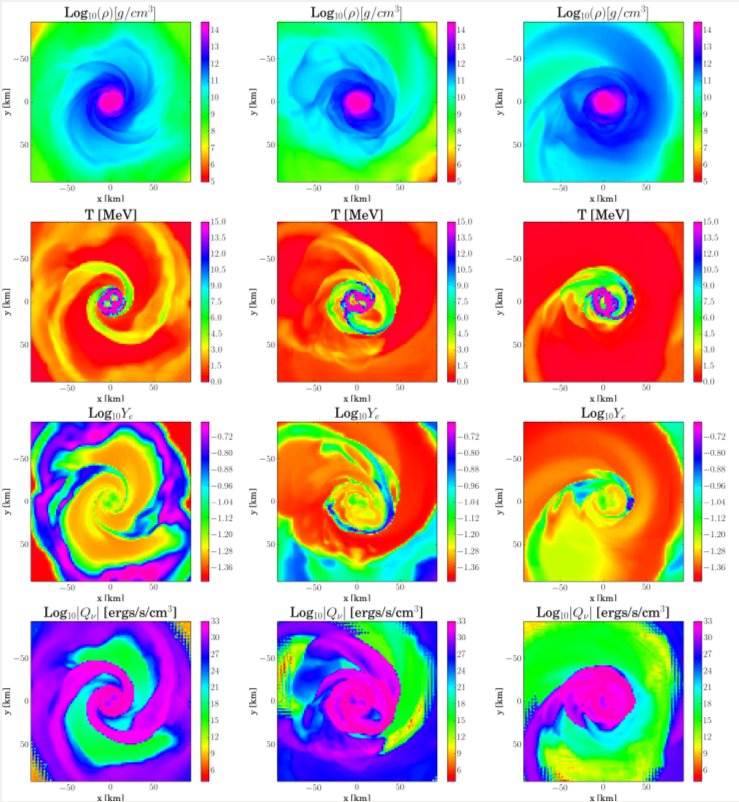}
\caption{Density, temperature, electron fraction, and neutrino luminosity rate for
 the DD2 EoS on the $z=0$ plane at roughly $t=3$~ms after the merger.
Shown are the three mass ratios $q=1$ (left), $q=0.85$ (middle) and $q=0.76$ (right).
} 
\label{fig:eos_2dplots_zplane_DD2}
\end{figure}

We begin by considering all three EoS for the same mass ratio, $q=0.85$.
Fig.~\ref{fig:eos_2dplots_zplane_q085} displays various properties along
the equatorial plane ($z=0$), all roughly 3 milliseconds after merger.
The density variations are similar to that seen in the equal mass case (see Figure 9 of~\cite{2015PhRvD..92d4045P}); 
the SFHo remnant is more centrally condensed compared to the remnants of the stiffer EoS that are less compact. 
Such behavior is expected as the collision takes place deeper in the gravitational potential of the system which,
in turn, implies the remnant of the SFHo case is hotter than the other two. Also, while all three snapshots with different EoS
show wisps of electron rich material just outside the remnants, the SFHo has the most
significant of such regions. This is due to the larger matter temperatures in the SFHo remnant that drives decompression of the hot material to lower densities where positron capture on neutrons can effectively raise the electron fraction. For all the EoS, there is evidence of tidal tails with one spiral arm stronger than the other. This asymmetry arises from the unequal mass ratio of the binary.

To more cleanly understand the role of mass ratio, we fix the EoS to that with the intermediate
stiffness, DD2, and consider three mass ratios. Fig.~\ref{fig:eos_2dplots_zplane_DD2}
displays the same quantities as in Fig.~\ref{fig:eos_2dplots_zplane_q085} but instead
for mass ratios $q=\{ 1, 0.85, 0.76\}$.

The density panels show that a decreased mass ratio results in a similar looking remnant but with
more dispersed material outside it due to stronger tidal effects. The temperature looks roughly similar in
the star, although slightly higher in the tidal tails of the equal mass cases.
The electron fraction is noticeably lower in the decreased mass ratio simulations as is the neutrino emissivity. 
All the quantities display a strong spiral arm in the case of unequal masses, in contrast to the two 
spiral arms of the equal mass case.
As we discuss later however, these broad properties do not convey what portion of this material
is bound/unbound which is important in determining possible kilonova characteristics.

\begin{figure}[h]
\centering
\includegraphics[width=9.0cm,angle=0]{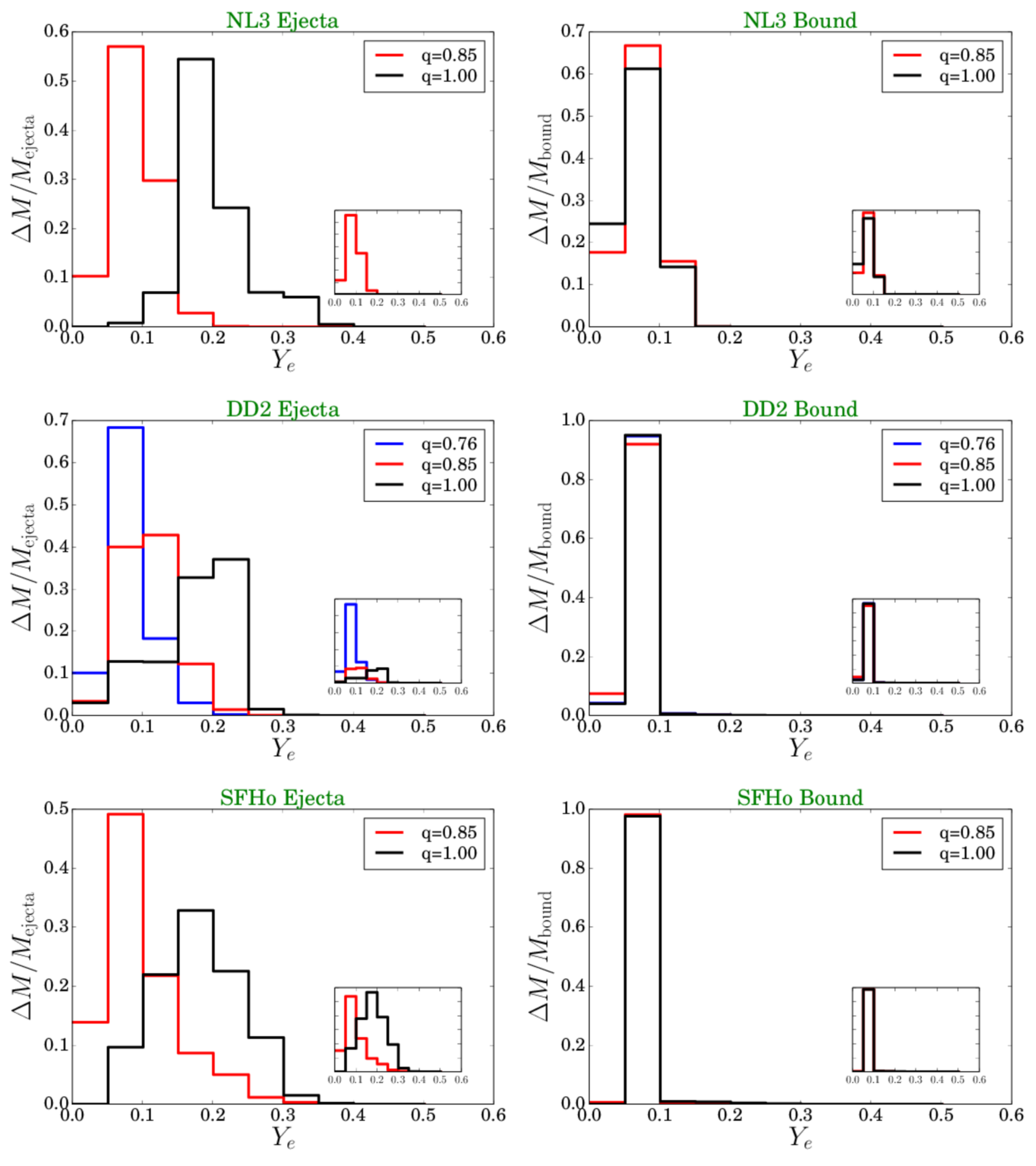}
\caption{ 
\textit{Distribution of the electron fraction for bound and unbound material.}
The fractional quantity of material is displayed for the different bins, and in
the insets the absolute quantities are shown.
For all three EoS, the bound material hardly changes with the mass ratio.
In contrast, the electron fraction generally decreases as the mass ratio
decreases.
}
\label{fig:histo_ye}
\end{figure}

\begin{figure}[h]
\centering
\includegraphics[width=9.0cm,angle=0]{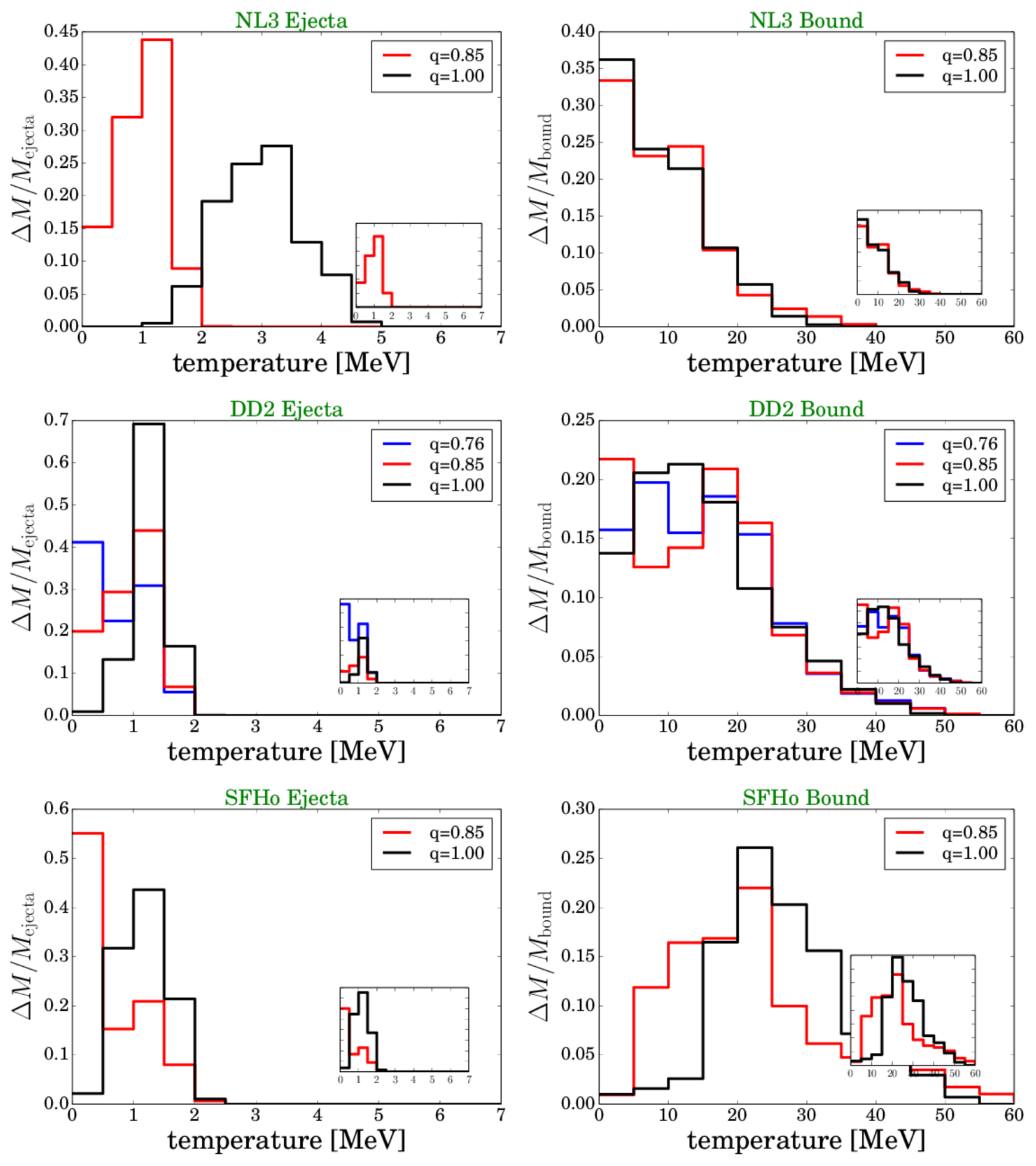}
\caption{ 
\textit{Temperature distribution of bound and unbound mass.}
As in Fig.~\ref{fig:histo_ye}, the insets show the absolute quantities in arbitrary
units. The ejecta is generally cold for any EoS and any mass ratio.
}
\label{fig:histo_temp}
\end{figure}

\begin{figure}[h]
\centering
\includegraphics[width=9.0cm,angle=0]{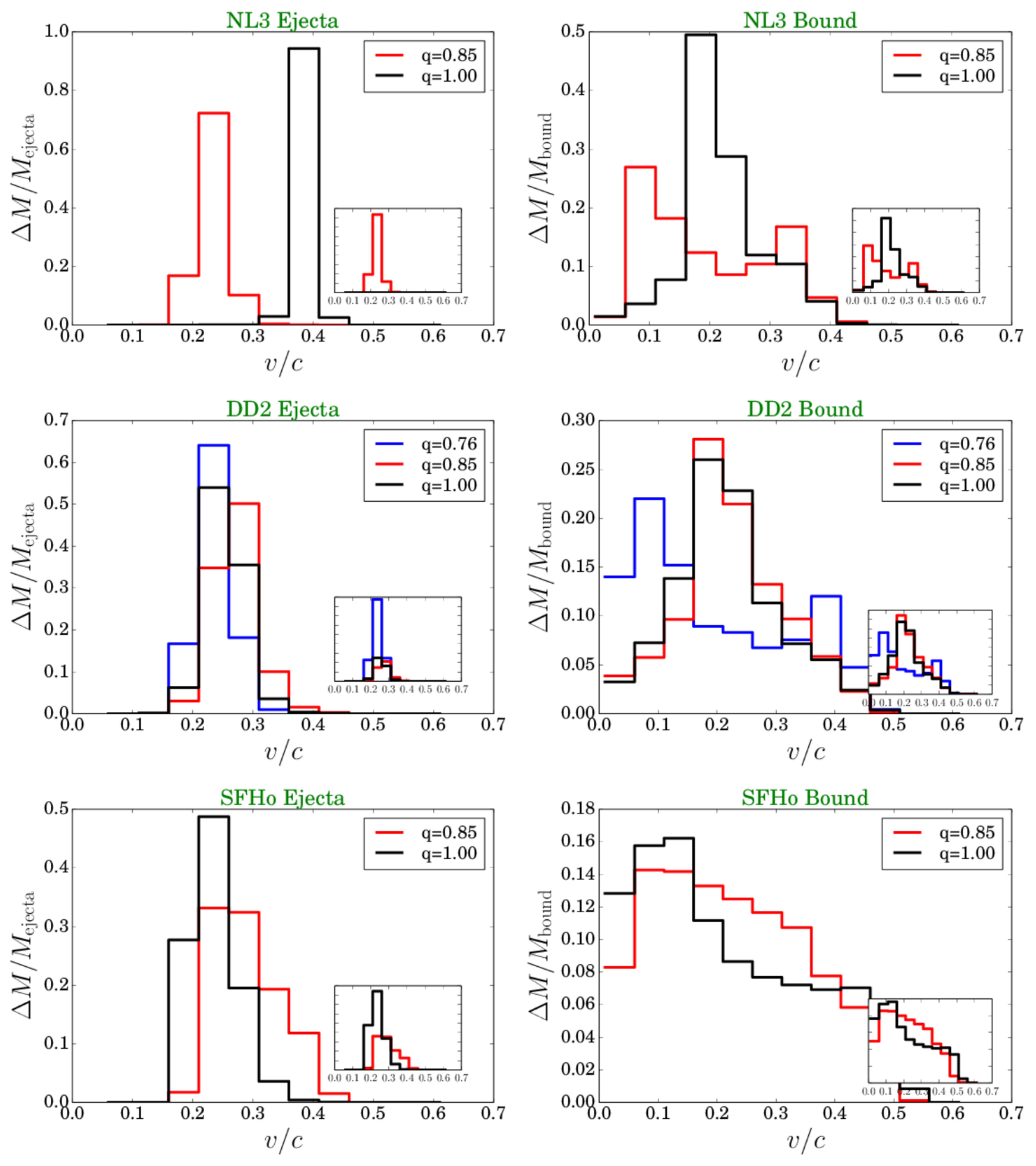}
\caption{ 
\textit{Velocity distribution of bound and unbound mass.}
As in Figs.~\ref{fig:histo_ye} and~\ref{fig:histo_temp}, the insets show the absolute quantities in arbitrary
units. 
}
\label{fig:histo_v}
\end{figure}

\begin{figure}[h]
\centering
\includegraphics[width=9.0cm,angle=0]{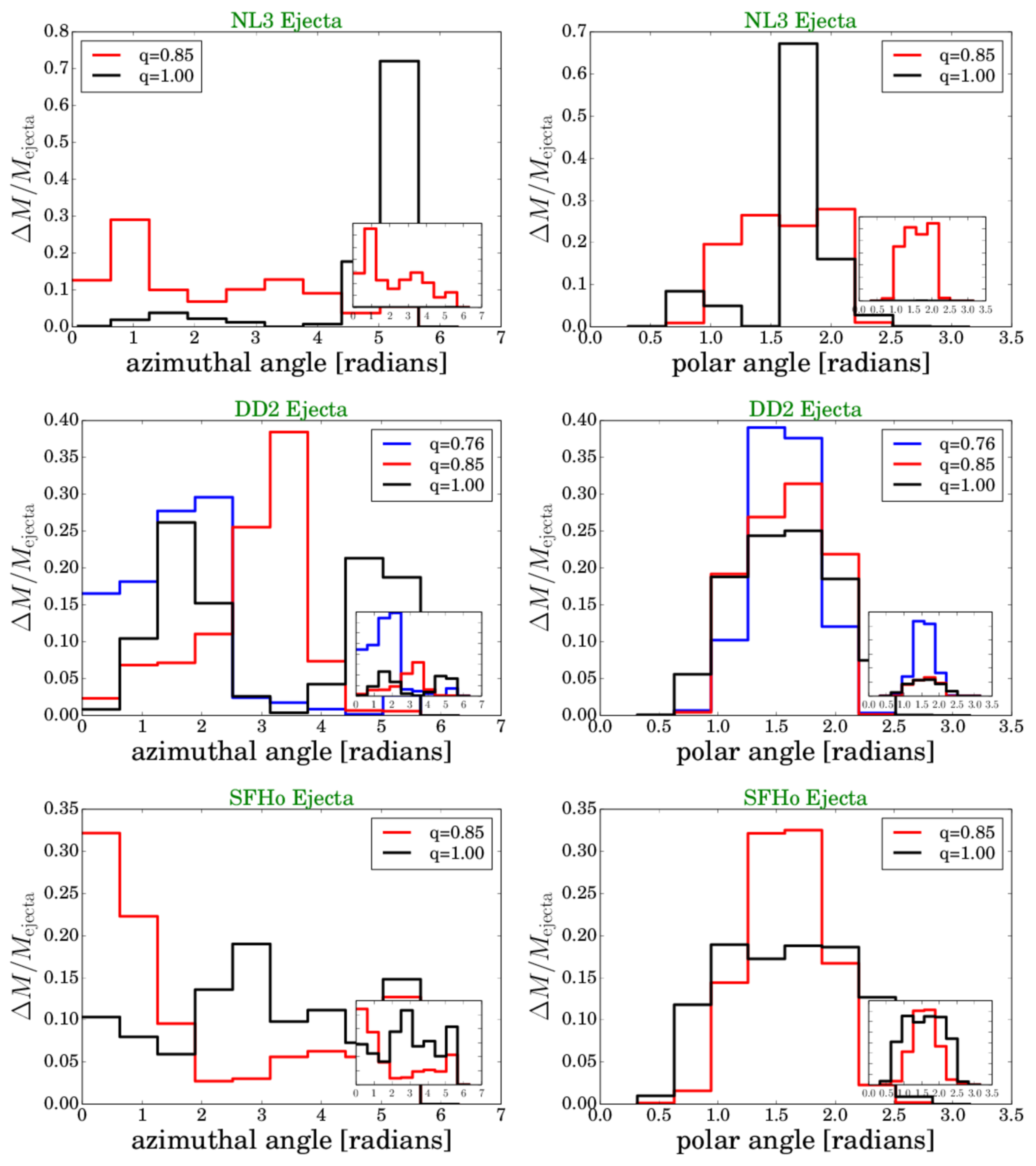}
\caption{ 
\textit{Azimuthal and polar angular distribution of unbound mass.}
As in Figs.~\ref{fig:histo_ye} and~\ref{fig:histo_temp}, the insets show the absolute quantities in arbitrary units. 
}
\label{fig:histo_angle}
\end{figure}

To this end, we integrate the characteristics
of both the bound and the unbound matter and plot them as histograms
(such calculations are described in Ref.~\cite{2015PhRvD..92d4045P}).
In particular, we display the electron fraction in Fig.~\ref{fig:histo_ye}, the temperature in Fig.~\ref{fig:histo_temp},
the velocity in Fig.~\ref{fig:histo_v} and the polar and azimuthal angles in Fig.~\ref{fig:histo_angle}.
Using these histograms, we 
present some observations and follow these with
a general discussion about these features.

Quite apparent in Fig.~\ref{fig:histo_ye} is that bound material is generally quite neutron
rich and independent of the mass ratio. However, the temperature distribution of this bound material does show some dependence on the mass ratio in Fig.~\ref{fig:histo_temp}. For the NL3 case, the dependence is minimal with the temperature distribution remaining generally the same among the mass ratios. The SFHo case shows the largest variation with mass ratio with the smaller mass ratio yielding a generally cooler merger remnant.

The ejecta, on the other hand, becomes increasingly neutron rich as the mass ratio departs from unity. Also apparent (in the insets) is that the amount of unbound material generally increases when the mass ratio decreases. We quantitatively summarize the amount of ejecta mass in Table~\ref{table:equal_mass}. The NL3 ejecta for the equal 
mass case is minuscule but significant for $q=0.85$. The DD2 ejecta increases much less dramatically, but the SFHo breaks with this
trend, albeit only moderately. The temperature of the ejecta is quite cool ($<5$ MeV) for every EoS and mass ratio,
as displayed in Fig.~\ref{fig:histo_temp}. The observed trend indicates that for lower values of $q$ the ejected material is cooler, being dominated by neutron-rich, tidal tail material. 

Another interesting observation comes from the angular distribution of the ejected
material. Fig.~\ref{fig:histo_angle} shows the azimuthal and polar angle distributions of unbound material. The histograms of
polar angle do not show significant qualitative differences between
the different mass ratio cases other than one observed trend. For low mass ratios the ejecta is more confined towards the equatorial region, giving quantitatively steeper distributions. The azimuthal distribution 
is more isotropic in the equal mass case (two spiral arms) but more anisotropic in the unequal mass cases (one spiral arm). 
As expected, the anisotropy grows as the mass ratio decreases from unity.

These observations suggest that a smaller mass ratio results in a less violent collision
because the less massive star disrupts earlier and further out of the gravitational well than in the equal mass case. Only the
NL3 bound material has essentially the same temperature for the different mass ratios. This EoS, however, has the largest stars
generally, and so its equal mass case is already fairly cool. In contrast, the SFHo case has noticeably cooler bound and ejected material.

In general terms, the unbound matter (or ejecta) can be sourced either
by tidal disruption during the merger or by thermal pressure and other interactions (e.g. shock heating) 
[On longer timescales --not studied here-- an extra source is provided by winds or outflows from an accretion disk~\cite{Fernandez:2014cna}].
Our work here concentrates mainly on the material produced by tidal interaction, and we observe that,
as the mass ratio decreases, the tidal disruption occurs earlier and further out. Consequently, there is more tidal-ejecta 
mass for small mass ratios.  Furthermore, the temperature of this ejecta is lower, which inhibits positron production and capture on neutrons. Thus,
the electron fraction is much closer to that of the original neutron star material.
Only the SFHo retains essentially
the same amount of ejected material with changes to the mass ratio because of a presumed balance of a competing effect; namely that
the equal mass case was quite violent resulting in significant interaction ejecta (with higher values of $Y_e$ and $T$) from shock heating. In the unequal mass case, the
violence decreases (decreasing the interaction ejecta), but the early disruption increases the tidal ejecta.  This explanation is consistent with the temperature and $Y_e$ distributions of ejecta mass discussed above and shown in Figs.~\ref{fig:histo_ye} and~\ref{fig:histo_temp}.  

The current work is so far the only one of which we are aware studying unequal binaries employing realistic microphysical
EoS; thus there is a lack of studies with which to compare, for instance, the amount of ejected material. 
As we did for the case of gravitational wave characteristics, we can compare our results
to those obtained employing piecewise polytropic EoS. To do so we focus on cases studied using
the so-called  MS1, H4, SLy piece-wise polytropes from~\cite{2015PhRvD..91l4041D} which have similar
mass ratios and masses to those used here for the NL3, DD2, SFHo EoS. 
The results are in reasonable agreement with some of the cases, with
typical ejected masses of $10^{-4}-10^{-3} M_{\odot}$ (although they 
find even $10^{-2} M_{\odot}$ for their softest EoS [SLy]). It is not
clear if these differences come only from the EoS choice. Nevertheless the
overall trend is the same, namely that \emph{a)} softer EoS have larger ejecta
and \emph{b)} unequal mass cases, for a given EoS, have larger ejecta masses for stiff EoS (DD2 and NL3 in our case, MS1 and H4 in the case of~\cite{2015PhRvD..91l4041D}) and lower ejecta masses for soft EoS (SFHo in our case, SLy in~\cite{2015PhRvD..91l4041D}). 

The ejecta values obtained here allow us to estimate the lag-times (with respect to merger) and luminosities
corresponding to each case. An important property of ejecta with such a low value of $Y_e$ is its ability
to produce elements with high  atomic mass number ($A\ge 120$)~\cite{1974ApJ...192L.145L,1989ApJ...343L..37M,1989Natur.340..126E}.
As argued in~\cite{Barnes:2013wka}, due to the increased
opacity as a result of lanthanides contributions, the expected
emission from thermal decay of r-process elements  would be
dimmer/redder and delayed with respect to typical supernova expectations.
Estimates in~\cite{Barnes:2013wka} (slightly re-arranged as in~\cite{East:2015vix}) argue that the rise time of the corresponding
kilonova emission $t^k_\mathrm{peak}$ 
 (from the time of the merger) is given by
\begin{equation}
t^{k}_\mathrm{peak}                     \approx 0.25 \,\mathrm{days} \left[\frac{M_\mathrm{eject}}{10^{-2} M_{\odot}} \right]^{1/2} \left[\frac{v}{0.3c}\right]^{-1/2} 
\label{eq:rprocesspeak}
\end{equation} 
with a  peak luminosity of
\begin{equation}
L \approx 2 \times 10^{41} \mathrm{erg/s} \left[\frac{M_\mathrm{eject}}{10^{-2} M_{\odot}} \right]^{1/2} \left[\frac{v}{0.3c}\right]^{1/2}.
\label{eq:rprocesslum}
\end{equation}
The values obtained for the cases studied are summarized in Table \ref{ejecta_Land_t}.

\begin{table*}\centering

\begin{ruledtabular}
\begin{tabular}{lllllllll}
EoS & $q$ &  $L[10^{40}$erg/s]  &  $t^{k}_\mathrm{peak}                        $ [days] & $M_{\rm eject} [10^{-3} M_{\odot}]$ & $v/c$ & $E_{\rm kin}[10^{50} \mathrm{ergs}]$ & $t_{\rm peak}$ [yr] &
$F(1\,\mathrm{GHz})$ [mJy] \\ \hline
NL3  & 1.0  &  0.9 & 0.008 & 0.015 & 0.45 & 0.01 & 0.31 & $1.8 \times 10^{-3}$ \\ 
NL3  & 0.85 &  8.8 & 0.13 & 2.3 & 0.25 & 1.22 & 4.0 & $4.4 \times 10^{-2}$ \\ 
DD2  & 1.0  &  4.1 & 0.05 & 0.43 & 0.3 & 0.31 & 1.9 & $1.9  \times 10^{-2}$\\ 
DD2  & 0.85 &  4.1 & 0.05 & 0.42 & 0.3 & 0.29 & 1.8 & $1.7  \times 10^{-2}$\\ 
DD2  & 0.76 &  7.2 & 0.09 & 1.3 & 0.3 & 0.76 & 2.5 & $4.6  \times 10^{-2}$ \\ 
SFHo & 1.0  & 10.6 & 0.16 & 3.4 & 0.25 & 1.8 & 4.6 & $6.5  \times 10^{-2}$\\ 
SFHo & 0.85 &  8.6 & 0.13 & 2.2 & 0.25 & 1.8 & 4.6 & $6.5  \times 10^{-2}$ \\
\end{tabular}
\end{ruledtabular}

\caption{Estimated properties of electromagnetic (EM) signals after the merger. 
The rise-time $t^k_{\rm peak}$ and luminosity $L$ of
EM radiation from a potential kilonova are calculated from 
Eqs.~\eqref{eq:rprocesspeak} and~\eqref{eq:rprocesslum}. 
We used the middle value of the bin corresponding to the peak velocity 
(see Fig~\ref{fig:histo_v}), $v/c$, in calculating these estimates.
The lag time $t_{\rm peak}$ and flux density $F$, estimated from the collision
of the ejecta with the interstellar medium, are calculated from 
Eqs.~\eqref{eq:collisionpeak} and~\eqref{eq:collisionflux}.
}
\label{ejecta_Land_t}
\end{table*}

In addition to the  electromagnetic signals produced by the decay of radioactive elements, another
transient has been proposed as a result from the collision of the ejected material
with the interstellar medium~\cite{Nakar:2011cw}. This interaction would produce
a radio emission that evolves more slowly. Indeed, the estimated timescales and brightness
are expected to obey~\cite{Nakar:2011cw} (employing the re-arranged version of~\cite{East:2015vix})
\begin{equation}
t_{\rm peak} \approx 6 \,\mathrm{yr} \left[ \frac{E_{\rm{kin}}}{10^{51} \mathrm{erg}} \right]^{1/3} \left[ \frac{n_0}{0.1 \,\mathrm{cm}^{-3}}\right]^{-1/3}
\left[\frac{v}{0.3c} \right]^{-5/3}
\label{eq:collisionpeak}
\end{equation}
with a  flux density
\begin{eqnarray}
F({\nu_\mathrm{obs}})  &\approx& 0.6 \,\mathrm{mJy} \left[ \frac{E_{\rm{kin}}}{10^{51} \mathrm{erg}} \right]  \left[ \frac{n_0}{0.1 \,\mathrm{cm}^{-3}}\right]^{7/8}
\left[\frac{v}{0.3c} \right]^{11/4} \nonumber \\
 & & \left[\frac{\nu_\mathrm{obs}}{1\,  \mathrm{GHz}} \right]^{-3/4} \left[\frac{d}{100\,  \mathrm{Mpc}} \right]^{-2}.
\label{eq:collisionflux}
\end{eqnarray}
Table~\ref{ejecta_Land_t} illustrates the values obtained for $t_{\rm peak}$ and $F({\nu_\mathrm{obs}})$ (for a representative value 
of $\nu_\mathrm{obs}= 1 \,\mathrm{GHz}$) assuming a density of $n_0 = 0.1  \,\mathrm{cm}^{-3}$ and a distance of $d=100\,\mathrm{Mpc}$. The
values of $E_{\rm{kin}}$ have been calculated for the unbound material as the total non-gravitational energy minus the rest mass and the internal energies
~\cite{2013PhRvD..87b4001H}.

\subsection{Neutrino Emission}
Finally we turn our attention to the neutrinos emitted by the system. 
We estimate the luminosity via a leakage scheme~\cite{Ruffert:1995fs,Rosswog:2003rv,O'Connor:2009vw}
and determine the thermodynamic properties of the neutrinospheres via a ray-tracing formalism~\cite{caballero:09}. The algorithms
we employ to obtain both estimates have been described in our previous
papers~\cite{Neilsen:2014hha} and~\cite{2015PhRvD..92d4045P}, respectively.

Figs.~\ref{fig:eos_2dplots_zplane_q085} and~\ref{fig:eos_2dplots_zplane_DD2}
illustrate both the spatial extent of the temperature and of the neutrino emissivities in the equatorial plane at $\sim$3\,ms after merger for the $q=0.85$ mass ratio simulations across all EoS (NL3, DD2, SFHo),
and for the DD2 EoS simulations across the three mass ratios considered ($q=1.0$, 0.85, and 0.76), respectively.
The first figure indicates that softer EoS will give rise
to higher neutrino luminosities which is consistent
with observations in the equal mass case~\cite{2015PhRvD..92d4045P}.
The higher luminosities are natural consequences of both the increased temperature and the increased amount of shock
heating for the softer EoS. The higher temperatures also imply that the neutrino average energies will be higher.

\begin{figure}[h]
\centering
\includegraphics[width=0.35\textwidth,angle=-90]{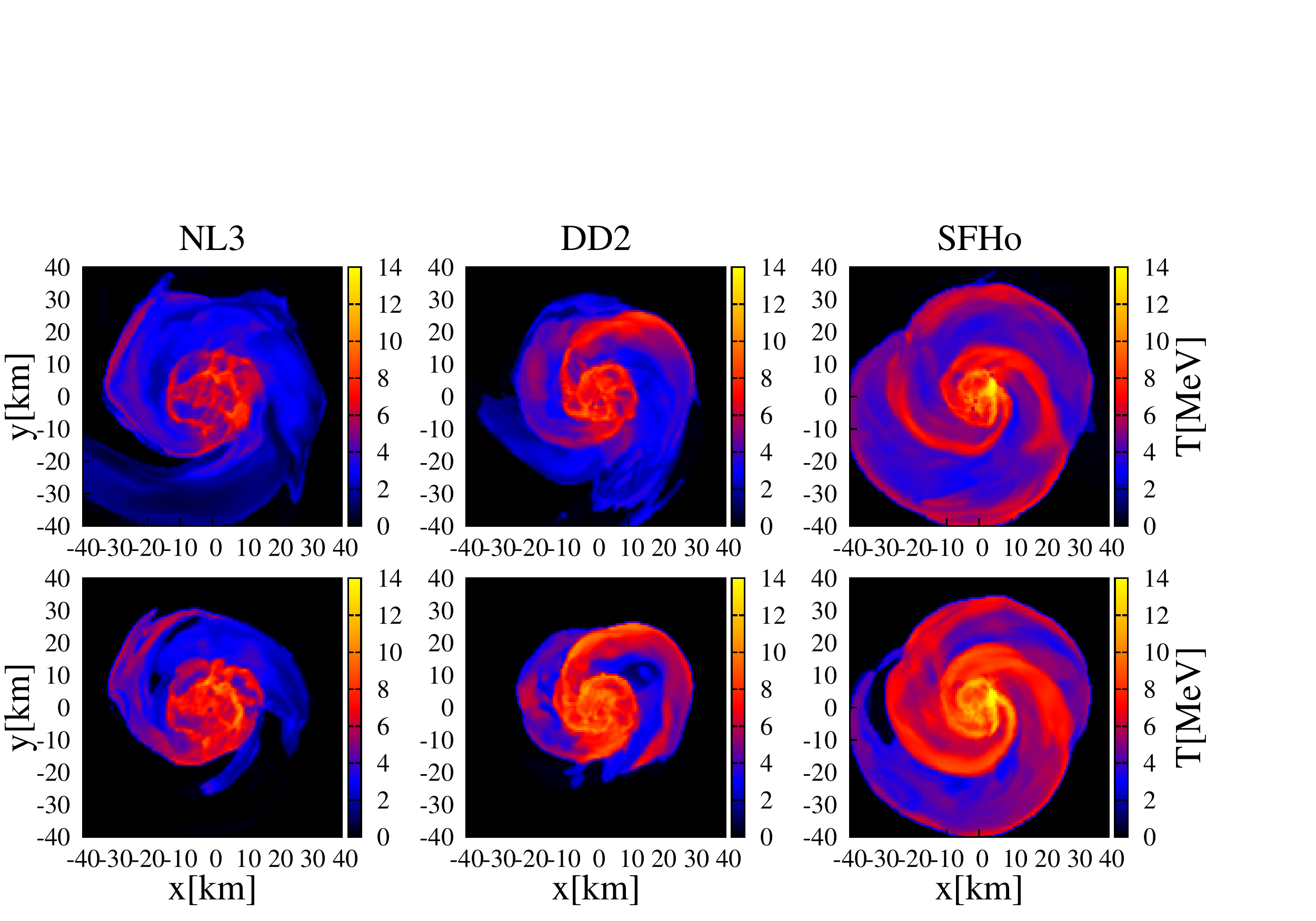}
\caption{ 
Electron neutrino (top) and electron antineutrino (bottom) surfaces as seen from the $z$-axis
for the three different EoS considered in this work
and a mass ratio $q=0.85$, $\sim$3 ms after the
merger.
}
\label{fig:EOS_neutrinos_q085}
\end{figure}

These observations and conclusions are also supported by the results
of our ray-tracing shown in Fig.~\ref{fig:EOS_neutrinos_q085}. This
figure shows the spatial distribution of the temperature of the
electron neutrinosphere and electron antineutrinosphere for the
$q=0.85$ mass ratio simulations using three different EoS at
$\sim$3\,ms after merger. They show higher peak matter temperatures at
the electron (anti)neutrinosphere for the SFHo EoS, locally as high as
$\sim$16\,MeV ($\sim$18\,MeV), when compared to the NL3 EoS where the
peak temperatures at the (anti)neutrinosphere reach $\sim$12\,MeV
($\sim$14\,MeV).  Our ray-tracing gives average electron neutrino
(antineutrino) energies (as measured at infinity) for these same
snapshots of $\sim$12.6\,MeV ($\sim$15.6\,MeV), $\sim$15.1\,MeV
($\sim$18.1\,MeV), and $\sim$15.3\,MeV ($\sim$17.8\,MeV), for the NL3,
DD2, and SFHo EoS, respectively. These average energies are also
summarized in Table~\ref{tab:qenergies}.

\begin{figure}[h]
\centering
\includegraphics[width=0.35\textwidth,angle=-90]{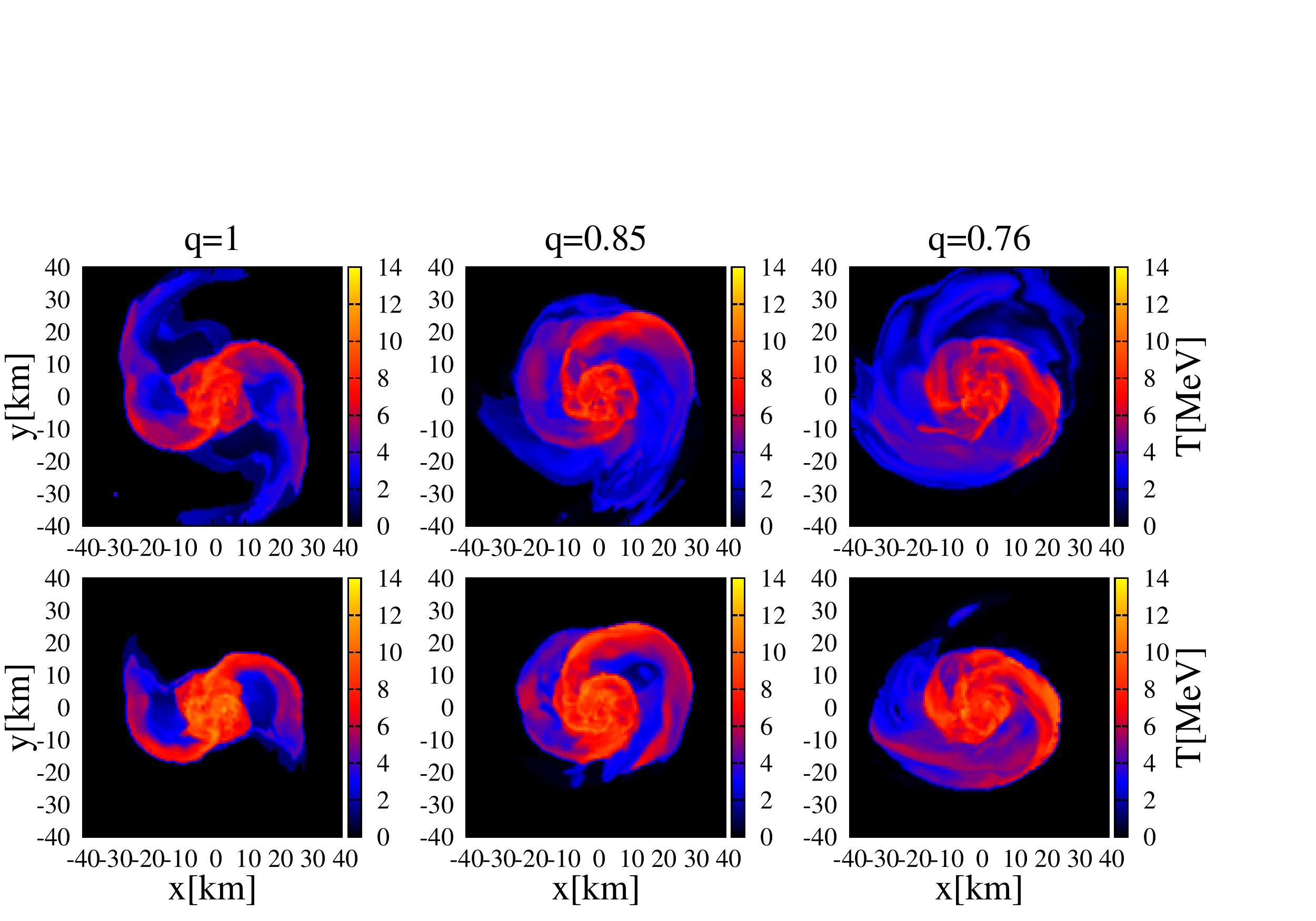}
\caption{ 
Electron neutrino (top) and electron antineutrino (bottom) surfaces as seen from  the $z$-axis 
for the DD2 EoS and different mass ratios $q$ at $t=3$~ms after the merger.
}
\label{fig:DD2_qs}
\end{figure}

We next consider the impact of the mass ratio on the properties of the
emitted neutrinos. Fig.~\ref{fig:DD2_qs} shows the neutrino surfaces $\sim$3\,ms
after merger for the three mass ratios considered for the
DD2 EoS. The maximum temperatures achieved at the neutrinospheres
are $\sim$14\,MeV and correspond to the equal mass case. These maximum
temperatures decrease to $\sim$12\,MeV for the smaller mass ratios. As
$q$ decreases, the matter distribution becomes more dispersed, a
consequence of stronger tidal effects, and this effect is reflected in
more dispersed neutrino surfaces that, in addition to reaching smaller
maximum temperatures, display a more uniform temperature. This more
dispersed, but uniform, behavior can also be seen close to the core in
the equatorial slices of Fig.~\ref{fig:eos_2dplots_zplane_DD2}. Our
neutrino analysis shows that for the NL3 EoS there is a drop of
$\sim$2--3\,MeV in the average neutrino energy emitted from the merger
remnants when the mass ratio goes from $q=1$ to $q=0.85$. For the DD2
EoS, no significant change or trend is seen in the neutrino average
energies for these three particular snapshots. 

The impact of the mass ratio on the neutrino emission properties of
the post merger evolution appears to be stronger for the SFHo EoS
simulations. Recall the observation in the temperature and $Y_e$
distributions (Figs.~\ref{fig:histo_temp} and~\ref{fig:histo_ye}) of
the ejected mass and bound mass from the previous section. For the
soft SFHo EoS, when comparing the $q=1$ to the $q=0.85$ mass ratio
simulation, we saw a transition from shock heated--or
interaction--ejecta to tidal ejecta and overall lower
temperatures. Our neutrino analysis suggests that this transition has a
direct and significant impact on the neutrino quantities; we observe
significantly lower average neutrino energies ($\sim$6-8\,MeV; see
Table~\ref{tab:qenergies}) and consequently predict a lower rate of
neutrino detection for close by mergers.  This analysis would indicate
that observations of a few systems with different values of $q$
(extracted from gravitational wave measurements of the individual NS
masses) could hint at properties of the underlying EoS (see e.g. \cite{lrr-2006-3}).

We caution however that these observation are drawn from a detailed
analysis at particular moments of time after merger in different
simulations. These disks are highly perturbed and are undergoing rapid
changes.  It is important to explore this possibility more fully in
the future by employing actual neutrino transport in order to obtain a
better handle on the neutrino energy and luminosity evolution as well
as to capture the effects of neutrino absorption.

\begin{figure}[h]
\centering
\includegraphics[width=9.2cm,angle=0,clip]{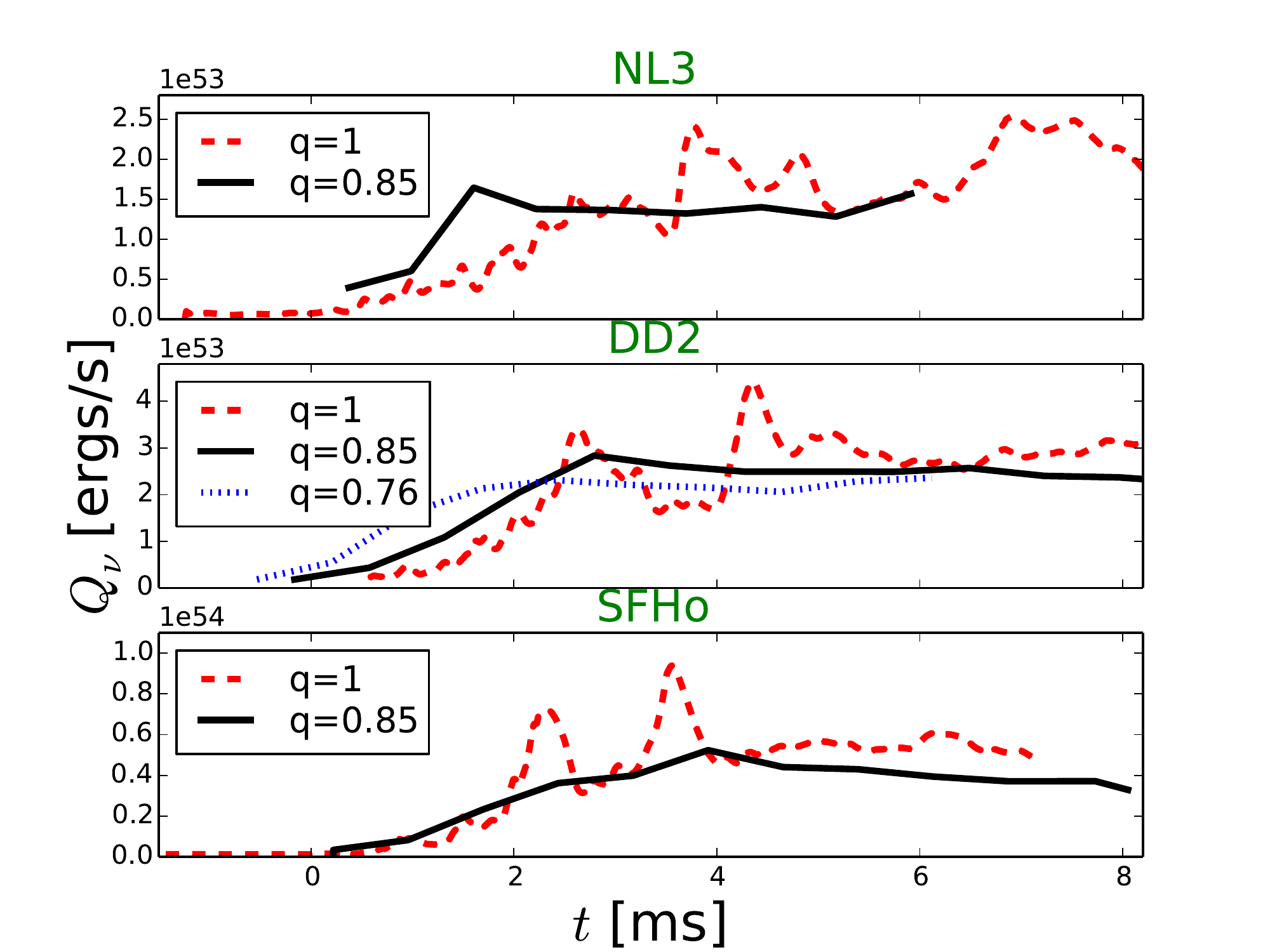}
\caption{Total neutrino luminosity from the BNS merger for the
  different mass ratios for each EoS. Shown are the NL3 EoS (top), DD2
  EoS (middle), and SFHo (bottom).  Note that the vertical scales
  differ in the three plots, with NL3 having the smallest range and
  SFHo the largest. Recall that mergers of unequal masses ($q<1$)
  occur early, by up to $\sim$1\,ms for the NL3 and DD2 EoS.}
\label{fig:eos_neutrinos}
\end{figure}

We end our discussion of the neutrino properties of the merger
remnants by exploring the evolution of the neutrino luminosity.
Fig.~\ref{fig:eos_neutrinos} shows the total neutrino luminosity as a
function of time for all cases considered. Since neutrino production
is strongly dependent on the temperatures achieved during and after
the merger, a higher emission is expected as the EoS softens. 
Our simulations follow this trend (note the different scales in
Fig.~\ref{fig:eos_neutrinos} for the different EoS).

As discussed throughout this paper, the merger yields
a less violent collision for a given EoS as the
mass ratio is decreased.  This
observation is also apparent in the neutrino luminosities. The equal
mass cases show various episodes of higher neutrino luminosity (and
neutrino average energy; see Table~\ref{tab:qenergies}) tied to the
decompression stage that follows both the bounce (i.e. intensity
maxima) and the occurrence of periods with higher relative intensity in
gravitational waves.  Our results suggest that unequal mass cases display a
less oscillatory total neutrino luminosity. Although our sampling
frequency of the this luminosity is less than it was for the equal
mass cases, we see no suggestions of the quite apparent oscillations
of the q=1 cases. Furthermore, physically one would naturally expect
that the oscillations would be more apparent in the symmetric, equal
mass case as the bar-like remnant oscillated about a centrally
condensed state.
Once again, we stress the importance of a
more detailed computation including neutrino transport to confirm the
conclusions reached here.

The very rare observation of neutrinos from a close-by (galactic)
neutron star merger would greatly complement the GW signature and
reveal important information regarding the nuclear EoS. In combination
with the electron antineutrino luminosities from our leakage scheme,
we can use the average energies computed from our ray-tracing to
estimate detection rates in a SuperKamiokande-like water-Cherenkov
detector.  These numbers are summarized in
Table~\ref{tab:qenergies}. In addition to the three $q=0.85$ mass
ratio simulations, we also include in the table the results from the
$q=1$ simulations (with all three EoS) and the $q=0.76$ simulation
performed with the DD2 EoS.  We note that the change in definition of
merger time between this work and that of~\cite{2015PhRvD..92d4045P}
causes slight differences in the value of $t$ for the $q=1$
snapshots~\footnote{In Ref.~\cite{2015PhRvD..92d4045P}, the merger time used
for the neutrino calculations measured time with respect to a merger time calculated
via the peak GW amplitude instead of the time of first contact.}.
 To estimate the neutrino detection rates presented in
Table~\ref{tab:qenergies}, we have made use of the following
approximation
\begin{equation}
R_\nu \sim \frac{21.05}{\mathrm{ms}}\left[\frac{M_\mathrm{water}}{32\,\mathrm{kT}}\right]\left[\frac{L_{\bar{\nu}_e}}{10^{53}\,
  \mathrm{erg}/ \mathrm{s}}\right]\left[\frac{\langle E_{\bar{\nu}_e} \rangle}{15\,\mathrm{MeV}}\right]\left[\frac{10\,\mathrm{kpc}}{D}\right]^2\,,\label{eq:detectionrate}
\end{equation}
where $L_{\bar{\nu}_e}$ and $\langle E_{\bar{\nu}_e} \rangle$ are the
electron antineutrino luminosity and average energy as listed in
Table~\ref{tab:qenergies}. The details of this expression and the
tracing technique are discussed in \cite{2015PhRvD..92d4045P}. Our
calculations suggest that discriminating the mass ratio based on
neutrino detection alone is unlikely. However such a detection,
especially if coincident or correlated with GWs, could shed light on
the dynamics of stellar material near, and soon after, the point of merger.

\begin{table}
\begin{center}
\begin{ruledtabular}
\begin{tabular}{ccccccc}
 EoS& $q$& $t$   & $\langle E_{\bar\nu_e}\rangle$ &$\langle E_{\nu_e}\rangle$ &$L_{\bar{\nu}_e}$   & $R_\nu$\\
    &    & [ms]  & $[$MeV$]$                      &$[$MeV$]$                  &$[10^{53}$ erg/s$]$ & $[$\#/ms$]$\\
  \hline
NL3  & 1.0  & 3.4 & 18.5 (22.4) & 15.2 (18.3) & 0.7 & 18  \\
NL3  & 0.85 & 3.0 & 15.6 (18.7) & 12.6 (15.1) & 0.8 & 18  \\ \hline
DD2  & 1.0  & 3.3 & 18.3 (22.1) & 14.6 (17.4) & 1.1 & 28  \\
DD2  & 0.85 & 2.8 & 18.1 (21.7) & 15.1 (18.0) & 1.0 & 25  \\
DD2  & 0.76 & 2.4 & 19.7 (23.9) & 14.8 (17.9) & 1.3 & 36  \\ \hline
SFHo & 1.0  & 3.5 & 24.6 (29.7) & 23.5 (28.3) & 3.5 & 121 \\
SFHo & 0.85 & 3.9 & 17.8 (21.3) & 15.3 (17.9) & 2.0 & 50  \\
\end{tabular}
\end{ruledtabular}
\
\caption{Neutrino properties at $\sim$3\,ms after the BNS merger for
  each EoS and mass ratio studied.  The average energies $\langle
  E_{\bar\nu_e}\rangle$ and $\langle E_{\nu_e}\rangle$ are computed
  with (without) the gravitational redshift from our ray-tracing
  scheme while the electron antineutrino luminosity for the
  corresponding time is predicted from the leakage scheme.  Also
  included is the estimated instantaneous detection rate in a
  SuperKamiokande-like water Cherenkov detector for a merger at
  10\,kpc from Earth, Eq.~\ref{eq:detectionrate}. The times are given
  with respect to $t_\mathrm{merger}$, which is defined for each EoS
  as the time when the $q=1$ mass ratio binary reaches first
  contact.}
\label{tab:qenergies}
\end{center}
\end{table}

\section{Conclusions}
\label{sec:conclusions}
This work furthers our studies of the rich phenomenology of binary mergers,
concentrating on unequal mass neutron star binaries with realistic equations 
of state and neutrino cooling. 
We focused on gravitational wave characteristics, ejecta and neutrino emission
from the system and, in particular, identified features that
can help extract the nuclear EoS from observations. As discussed, 
this information is encoded in 
observable signatures of gravity waves, electromagnetic radiation, and neutrinos, but
extracting this information with a detection in only one of these channels might prove
quite difficult. However, now that we have entered the era of GW detections, the
possibility of collecting multiple merger events in different bands offers the promise
of a deep understanding of the dynamics and EoS of these BNS systems.

Recent studies of BNS mergers have shown that the dominant frequencies of the post-merger GW
signals reveal details of the EoS. However, the region of highest sensitivity for current GW detectors
does not extend up to these merger frequencies, which limits our ability
to extract EoS characteristics due to tidal effects. Before merger, tidal effects only
become significant at the close separations achieved just before merger, and, again, these
frequencies are generally higher than the most sensitive bandwidth of GW detectors.

We have found that the peak frequency in the gravitational waveform 
can be estimated via a fit based on a straightforward
contact-frequency criteria which, in turn, relates directly to the EoS by the stellar radii. 
If the merger results in a kilonova, the resulting electromagnetic counterpart encodes more information
about the nuclear EoS.
We studied the potential for equal mass binaries to produce a kilonova in recent work~\cite{2015PhRvD..92d4045P}
finding the necessary conditions only for the softest EoS studied.
However, as the mass ratio $q$ decreases, we find that all three EoS considered produce 
significant ejecta with high neutron richness capable of  producing a kilonova that peaks in the infrared
(due to the production of heavy elements $A>120$). 
For maximum effect therefore, electromagnetic observations of kilonovae afterglows from BNS need to be combined with gravitational wave
measurements of the individual stellar masses.
Neutrinos are also produced copiously by the merger and their energies appear to depend 
more strongly with the mass ratio for softer EoS, though further studies are required
to confirm and quantify this possibility. 
Provided the merger occurs close
enough to detect any neutrinos, their observation will provide
valuable information on the EoS and complement the gravitational wave signature.

With the recent opening of the gravitational wave window, observing compact binary mergers
through gravitational waves has become a reality. The combination of this new information with
that enabled by electromagnetic signals and astroparticles will enable  a deep understanding 
of such systems.

%
%
\vspace{0.5cm}

\begin{acknowledgments}
It is a pleasure to thank William East, Francois Foucart, Jonas Lippuner and 
Eric Poisson  for interesting discussions as well as our 
collaborators Eric Hirschmann, Patrick Motl, and Marcelo Ponce.
This work was supported by the NSF under grants PHY-1308621~(LIU),
PHY-0969811 \& PHY-1308727~(BYU), NASA's ATP program through grant NNX13AH01G,
NSERC through a Discovery Grant (to LL) and CIFAR (to LL).
CP acknowledges support from the Spanish Ministry of Education and
Science through a Ramon y Cajal grant and from the Spanish Ministry of
Economy and Competitiveness grant FPA2013-41042-P.
Additional support for this work was provided by
NASA through Hubble Fellowship grant \#51344.001-A (EO) awarded by the
Space Telescope Science Institute, which is operated by the
Association of Universities for Research in Astronomy, Inc., for NASA,
under contract NAS 5-26555.
Research at Perimeter Institute is supported through Industry Canada and 
by the Province of Ontario
through the Ministry of Research \& Innovation.  Computations were
performed at XSEDE and Scinet. 
\end{acknowledgments}

\bibliographystyle{utphys}

\providecommand{\href}[2]{#2}\begingroup\raggedright\endgroup

\end{document}